\documentclass[preprint,12pt]{elsarticle}

\usepackage{graphicx, latexsym,amssymb,amsmath, float, enumitem}
\usepackage{algorithm}
\newtheorem{theorem}{Theorem}

\newtheorem{observation}[theorem]{Observation}
\newtheorem{definition}[theorem]{Definition}
\newtheorem{lemma}[theorem]{Lemma}
\newtheorem{corollary}[theorem]{Corollary}

\newtheorem{property}[theorem]{Property}

\begin{document}

\begin{frontmatter}


\title{Towards a provably resilient scheme for \\graph-based watermarking}

\tnotetext[wg]{An extended abstract of this paper
was presented at the 
39th International Workshop on Graph Theoretic Concepts in Computer Science, WG 2013, 
and appeared in \emph{Lecture Notes in Computer Science} \textbf{8165} (2013), 50--63. }

\author[addr1,addr2]{Lucila M. S. Bento}
\ead{lucilabento@ppgi.ufrj.br}

\author[addr2]{Davidson Boccardo}
\ead{drboccardo@inmetro.gov.br}

\author[addr2]{Raphael C. S. Machado}
\ead{rcmachado@inmetro.gov.br}

\author[addr1]{Vin{\' i}cius G. Pereira de S{\' a}}
\ead{vigusmao@dcc.ufrj.br}

\author[addr1,addr2,addr3]{Jayme Luiz Szwarcfiter}
\ead{jayme@nce.ufrj.br}

\address[addr1]{Instituto de Matem\'atica, Universidade Federal do Rio de Janeiro, Rio de Janeiro, Brasil}
\address[addr2]{Instituto Nacional de Metrologia, Qualidade e Tecnologia, Rio de Janeiro, Brasil}
\address[addr3]{COPPE Sistemas, Universidade Federal do Rio de Janeiro, Rio de Janeiro, Brasil}

\begin{abstract}
Digital watermarks have been considered a promising way to fight software piracy.
Graph-based watermarking schemes encode authorship/ownership data as control-flow graph of dummy code.
In 2012, Chroni and Nikolopoulos developed an ingenious such scheme which 
was claimed to withstand attacks in the form of a single edge removal. We extend the work of those authors in various aspects. First, we give a formal characterization of the class of graphs generated by their encoding function. Then, we formulate a linear-time algorithm which recovers from ill-intentioned removals of $k \leq 2$ edges, therefore proving their claim. Furthermore, we provide a simpler decoding function and an algorithm to restore watermarks with an arbitrary number of missing edges whenever at all possible. By disclosing and improving upon the resilience of Chroni and Nikolopoulos's watermark, our results reinforce the interest in regarding it as a possible solution to numerous applications.
\end{abstract}



\end{frontmatter}


\section{Introduction}

The illegal reproduction of software has become a major concern for the industry. According to the Business Software Alliance, the commercial value of unlicensed software put into the world market in $2011$ totaled $63.4$ billion dollars~\cite{bsa}. To counter such practice, many promising methods have been devised, among which the idea 
of software watermarking.  

The use of paper watermarks to prevent counterfeiting dates back to the thirteenth century. 
Generally speaking, watermarks are unique identifiers 
embedded into
proprietary objects to 
enforce authenticity. 
In a digital object, particularly in a piece of software,
a watermark may act not only as a certificate of authorship, 
but also as a means of tracing the original owner of the object, therefore discouraging piracy.

The first
software watermark was proposed in $1996$ by 
Davidson and Myrhvold \cite{davidson}, 
while the first 
watermarking scheme to exploit concepts of Graph Theory was formulated by Venkatesan, Vazirani and Sinha~\cite{venkatesan} in~$2001$. 
Their technique, whereby 
an integer was encoded as a special digraph
disguised into the software's
control-flow graph, was later patented~\cite{patent}. 
Other original ideas, improvements and surveys on the available methods have been contributed by many authors ever since. See, for example,~\cite{collberg, collberg-2, collberg-3, hamilton, su, zhu, zhu2}. 

Willing to prevent the timely retrieval of the encoded identification data,
malicious agents may attempt to tamper with the watermark.
A watermark solution is therefore only as secure as it is 
able to resist attacks of various sorts. Naturally,
a lot of research has been put up lately towards developing more resilient solutions as well as strengthening existing ones. This paper pursues this latter goal.

We consider the 
graph-based
watermarking scheme introduced by Collberg, Kobourov, Carter and
Thomborson \cite{collberg}, and afterwards developed and improved
upon by Chroni and Nikolopoulos in a series of \mbox{papers~\cite{chroni, chroni-6, chroni-2, chroni-3, chroni-4}.}
These latter authors proposed a watermark graph 
belonging to a subclass of the 
reducible permutation graphs introduced by the former authors.
Though the
mechanics of encoding and decoding the proposed watermark is well
described in \cite{chroni-2}, such special subclass of reducible permutation graphs has
not been fully characterized. 
Moreover, not much was known thus far about
the resilience of Chroni and Nikolopoulos's graphs 
to malicious attacks, even though their ability to withstand single edge removals has been suggested without proof.




This paper is organized as follows. 
In Section~\ref{s:preliminaries}, we
present some preliminary concepts related to 
graph-based
software watermarking, 
including the most common forms of attacks.
In Section~\ref{s:nikolopoulos}, we recall the watermark from
Chroni and Nikolopoulos, and we state 
a number of structural properties, the proofs of which 
we delay until Section~\ref{s:proofs} for the sake of readability.
In Section~\ref{s:prg}, we define and characterize the 
family of canonical reducible permutation graphs, which correspond to the watermarks
produced by Chroni and Nikolopoulos's encoding function.
In Section~\ref{s:newalg}, we formulate linear-time algorithms 
to reconstruct the original digraph and recover the encoded data
even if 
two edges are missing. The proof of one of the central results in that section, namely Theorem~\ref{thm:alg1}, 
is somewhat involved, and we dedicate a whole section to it towards the end of the paper.
In Section~\ref{s:poly}, we 
propose a robust polynomial-time algorithm that, given
a watermark with 
\emph{whatever} number $k$ of missing edges, either recovers the encoded data 
or proves that the the watermark has become irremediably damaged.
Finally, Sections~\ref{s:proofs} and~\ref{s:proofthm} contain the postponed proofs 
for the properties stated in Section~\ref{s:nikolopoulos} and for Theorem~\ref{thm:alg1}, respectively. 
Section~\ref{s:conclusion} concludes the paper with our final
remarks.

Throughout the text, we let $V(G)$ and $E(G)$ respectively denote, as usual, the
vertex set and edge set of a given graph $G$. Also, we let
$N_G^+(v)$ and $N_G^-(v)$ be the sets of
out-neighbors and in-neighbors of vertex $v$ in $G$,
with $d_G^+(v)$ and $d_G^-(v)$ their respective sizes. If $J$ is a subset of either $V(G)$ or $E(G)$, then $G - J$ corresponds to the graph obtained from $G$ by the removal of $J$.

\section{Graph-based software watermarking}\label{s:preliminaries}

Software watermarking schemes provide the necessary means of embedding identification data---typically a copyright notice or a customer number---into a piece of software. 
We refer to the identification data as the \emph{identifier}, and we may regard it as an integer, for simplicity.
Watermarks are appropriate encodings of identifiers, and they can be broadly divided into
two categories: static and dynamic~\cite{collberg-1}. 
The former are embedded in the code, 
whereas the latter are embedded into a program's execution state at runtime.

A static, \emph{graph-based} watermarking scheme usually consists of four algorithms:
\begin{itemize}
\item an \emph{encoder}, which converts the identifier into a graph---the watermark;
\item a \emph{decoder}, which extracts the identifier from the watermark;
\item an \emph{embedder}, a function whose input parameters are the software itself (either the binary code or the source code in some programming language), the intended watermark and possibly some secret key, and whose output is a modified software containing the watermark; and 
\item an \emph{extractor}, which retrieves the watermark graph from the watermarked software.
\end{itemize}

To every computer program one can associate a directed graph representing the possible sequences of instructions (or, more precisely, of jump-free instruction blocks) during its execution. Such graph, called the \emph{control-flow graph} (CFG) of the software~\cite{cfg}, can be obtained by means of static analysis~\cite{nielson, static-analysis}. What the embedder does is basically to insert dummy code into the program so that the intended watermark graph shows up as an induced subgraph of the CFG. The position of the watermark graph within the CFG is often determined as a function of a secret key. Knowledgeable of the secret key, the extractor retrieves that subgraph, which is then passed along to the decoding algorithm. Among the existing tools for embedding/extracting graph-based watermarks, we cite~Collberg's SandMark project~\cite{sandmark}.
In this paper, we focus on the encoding/decoding algorithms described by Chroni and Nikolopoulos in~\cite{chroni-2}.

\paragraph{Attacks} 
Among the several different kinds of attacks against graph-based watermarks, we list the following:
\begin{itemize}
\item \emph{additive} attacks, in which other watermarks are inserted into the same object, generating ambiguity; 
\item \emph{subtractive} attacks, in which the watermark is removed altogether; and
\item \emph{distortive} attacks, in which the watermark is modified to confound the decoder.
\end{itemize}

Additive and subtractive attacks can be precluded to a great extent by techniques of cryptography and software diversity~\cite{diversity}. On the other hand, distortive attacks---also known as \emph{jamming} attacks---are more difficult to deal with and are arguably the most important attack model to be concerned about~\cite{venkatesan}. In some cases, the distortive attacker may even be able to reverse engineer the entire code and apply semantics-preserving modifications which modify the CFG structurally without affecting the software's functionalities.

\section{The watermark by Chroni and Nikolopoulos} \label{s:nikolopoulos}

We recall the encoding algorithm
described in~\cite{chroni-2}. The index of the first element in all considered sequences is $1$.

Let $\omega$ be a positive integer identifier,
and $n$ the size of the binary representation $B$ of $\omega$. Let also
$n_{0}$ and $n_{1}$ be the number of $0$'s and $1$'s, respectively, in $B$,
and let $f_{0}$ be the index 
of the leftmost $0$ in $B$.
The extended binary $B^*$ is obtained by concatenating $n$ digits $1$, followed by the one's complement 
of $B$ and by a single digit $0$.
We let $n^* = 2n+1$ denote the size of $B^*$, and
we define \mbox{$Z_0=(z_i^0)$}, \mbox{$i = 1, \ldots, n_1+1$}, as the ascending sequence of indexes of $0$'s in $B^*$,
and \mbox{$Z_1=(z_i^1)$}, \mbox{$i = 1, \ldots, n + n_0$}, as the ascending sequence of indexes of $1$'s in $B^*$.

Let $S$ be a sequence of integers. We denote by $S^R$ the sequence formed by the elements of $S$ in backward order.
If $S = (s_i)$, for $i = 1, \ldots, t$, 
and there is an integer $k \leq t$ such that the subsequence
consisting of the elements of $S$ with indexes less than or equal to $k$ is ascending,
and the subsequence consisting of the elements of $S$ with indexes greater than or equal to $k$ is descending,
then we say $S$ is \emph{bitonic}.
If all $t$ elements of a sequence $S$ are distinct and belong to $\{1, \ldots, t\}$, then $S$ is a \emph{permutation}.
If $S$ is a permutation of size $t$, and, for all $1 \leq i \leq t$, the equality $i = s_{s_i}$ holds, then we say $S$ is \emph{self-inverting}.
In this case, the unordered pair $(i, s_i)$ is called a \emph{$2$-cycle} of $S$, if $i \neq s_i$, and a \emph{$1$-cycle} of $S$, if $i = s_i$.
If $S_1,S_2$ are sequences (respectively, paths in a graph), we denote by $S_1||S_2$ the sequence (respectively, path) formed by the elements of $S_1$ followed by the elements of $S_2$.

Back to Chroni and Nikolopoulos's algorithm, we define $P_b = (b_i)$, with $i=1,\ldots,n^*$, as the bitonic permutation $Z_0||Z_1^R$.
Finally,
the self-inverting permutation $P_s = (s_i)$ is
obtained from $P_b$ as follows:
for $i = 1,\ldots,n^*$,
element $s_{b_i}$ is assigned value $b_{n^* - i + 1}$,
and element $s_{b_{n^* - i + 1}}$ is assigned value~$b_i$.
In other words, the $2$-cycles of $P_s$ correspond to
 the $n$ unordered pairs of distinct elements of $P_b$
that share the same minimum distance to one of the extremes of $P_b$, 
that is, the pairs $(p,q) = (b_i, b_{n^* - i + 1})$, for $i = 1, \ldots, n$.
Since the central index $i = n+1$ of $P_b$
is the solution of equation $n^* - i + 1 = i$, element $b_{n+1}$ --- and no other --- will constitute a $1$-cycle in $P_s$.
We refer to such element of $P_s$ as its \emph{fixed element}, and we let $f$ denote it.

The watermark generated by Chroni and Nikolopoulos's encoding algorithm~\cite{chroni-2} 
is a directed graph $G$ whose vertex set is $\{0, 1, \ldots, 2n+2\}$,
and whose edge set contains $4n+3$ edges, to wit: a \emph{path edge} $(u, u-1)$ for $u=1,\ldots,2n+2$, constituting a Hamiltonian path that will be unique in $G$,
and a \emph{tree edge}
from $u$ to $q(u)$, for $u=1,\ldots,n^*$, where $q(u)$ is defined as the vertex $v > u$ with the greatest index in $P_s$ to the left of $u$, if such $v$ exists, or $2n+2$ otherwise. 
The rationale behind the name \emph{tree edge} is the fact that such edges induce a spanning tree of $G \setminus \{0\}$.

\begin{figure}
    \centering
    \includegraphics[width=\textwidth]{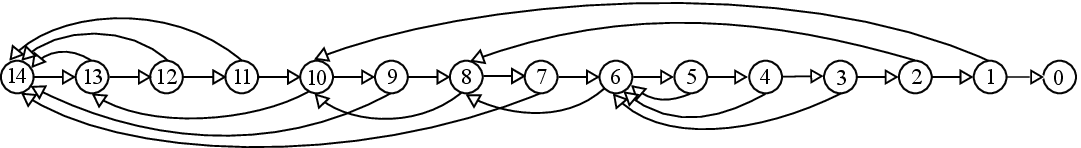}
    \caption{\label{f:graph43} Watermark for identifier $\omega = 43$.}
\end{figure}

Let us glance at an example. For \mbox{$\omega = 43$}, we have
\mbox{$B = 101011$},
\mbox{$n = 6$},
\mbox{$n_0 = 2$},
\mbox{$n_1 = 4$},
\mbox{$f_0 = 2$},
\mbox{$B^* = 1111110101000$},
\mbox{$n^* = 13$},
\mbox{$Z_0 = (7, 9, 11, 12, 13)$},
\mbox{$Z_1 = (1, 2, 3, 4, 5, 6, 8, 10)$},
$P_b = (7, 9, 11, 12, 13, 10, 8, 6, 5, 4, 3, 2, 1)$,
$P_s = (7, 9,\linebreak  11, 12, 13, 10, 1, 8, 2, 6, 3, 4, 5)$ and
\mbox{$f = 8$}. 
The watermark graph associated to $\omega$ presents, along with the path edges in the Hamiltonian path 
$14, 13, \ldots, 0$, 
the tree edges $(1,10), (2,8), (3,6), (4,6), (5,6), (6,8), (7,14), (8,10), (9,14), \linebreak (10,13), (11,14), 
(12,14)$ and $(13,14)$, as illustrated in Figure~\ref{f:graph43}.

In Section~\ref{s:prg}, we give a formal characterization of the class of such graphs, allowing for linear-time recognition and the formulation of efficient recovering algorithms against distortive attacks.

\subsection{Structural properties}\label{s:properties}

We now state a number of properties concerning 
the watermark from Chroni and Nikolopoulos
and the special permutations they are associated to. 
These properties, 
whose proofs are given in Section~\ref{s:proofs},
set the basis for the characterization of the class of canonical reducible permutation graphs, which is given in~Section~\ref{s:prg}, and
for the recovering procedures described in~Section~\ref{s:newalg}.

For all properties stated below, let $G$ be the watermark graph
associated to an identifier $\omega$ of size $n$,
and let $P_b$ and $P_s$ be, respectively, 
the bitonic and the self-inverting permutations 
dealt with during the construction of $G$.

\begin{property}\label{prop_end_pb}
For $1 \leq i \leq n$, the element $b_{n+i+1}$ in $P_b$ is equal to $n-i+1$, that is,
the $n$ rightmost elements in $P_b$, from right to left, 
are $1, \ldots, n$.
\end{property}

\begin{property}\label{prop_n_primeiros_elementos_maiores_que_n}
The elements whose indexes are $1,\ldots,n$ in $P_s$ are all greater than $n$.
\end{property}

\begin{property}\label{prop_fixo}
The fixed element $f$ satisfies
$f = n+f_0$, unless the identifier $\omega$ is equal to $2^k - 1$ for some integer $k$, whereupon $f = n^* = 2n+1$.
\end{property}

\begin{property}\label{prop_primeiros_elementos}
In self-inverting permutation $P_s$, elements indexed
$1, \ldots, \linebreak f-n-1$
are respectively equal to $n+1, n+2, \ldots, f-1$,
and elements indexed
$n+1, n+2, \ldots, f-1$
are respectively equal to $1, \ldots, f-n-1$.
\end{property}

\begin{property}\label{prop_elemento_central_1}
The first element in $P_s$ is $s_1 = n+1$, and the central element in $P_s$ is $s_{n+1} = 1$.
\end{property}

\begin{property}\label{prop_2n_plus_1}
If $f \neq n^*$, then
the index of element $n^*$ in $P_s$ is equal to $n_1 + 1$, and vice-versa.
If $f = n^*$, then the index of element $n^*$ in $P_s$ is also $n^*$.
\end{property}

\begin{property}\label{prop_inicio_bitonico}
The subsequence of $P_s$ consisting of elements indexed $1, \ldots, \linebreak n+1$ is bitonic.
\end{property}

\begin{property}\label{prop_uk2}
For $u \leq 2n$, 
$(u,2n+2)$ is a tree edge of watermark $G$ if, and only if,
$u-n$ is the index of a digit $1$ in the binary representation
$B$ of the identifier $\omega$ represented by~$G$.
\end{property}

\begin{property}\label{prop_uk}
If $(u,k)$ is a tree edge of watermark $G$, with $k \neq 2n+2$, then
\begin{enumerate}[label=(\roman*)]
\item element $k$ precedes $u$ in $P_s$; and
\item if $v$ is located somewhere between $k$ and $u$ in $P_s$, then $v < u$.
\end{enumerate}
\end{property}

\section{Canonical reducible permutation graphs}\label{s:prg}

This section is devoted to the characterization of the class of
canonical reducible permutation graphs. 
After describing some terminology and proving some preliminary results,
we define the class using purely graph-theoretical predicates.
Then, we show it corresponds exactly
to the set of watermarks  
produced by Chroni and Nikolopoulos's 
encoding algorithm~\cite{chroni-2}. 
Finally, we characterize it in a way 
that suits the design of 
a decoding algorithm which is simpler---for untampered with watermarks---and
able to recover from removals of $k \leq 2$ edges in linear time. 

A {\it reducible flow graph}~\cite{hecht, hecht-1, tarjan} is a 
directed
graph $G$ with source $s \in V(G)$, such that, for each cycle $C$
of $G$, every directed path from $s$ to $C$ reaches $C$ at the same
vertex. It is well known that a reducible flow graph has at most one
Hamiltonian cycle. 

\begin{definition}\label{def:slrfg}
A \emph{self-labeling reducible flow graph} 
is a directed graph $G$ such that
\begin{enumerate}[label=(\roman*)]
\item $G$ presents exactly one directed
Hamiltonian path $H$, hence there is a unique labeling function 
$\sigma~\colon V(G)
\to \{0, 1, \ldots, |V(G)|-1\}$ of the vertices of $G$ such that the
order of the labels along $H$ is precisely $|V(G)|,
|V(G)|-1, \dots, 0$; and,
\item considering the labeling $\sigma$ as in the previous item, 
         \mbox{$N_G^+(0) = \emptyset$}, \linebreak
         \mbox{$N_G^-(0) = \{1\}$}, 
         \mbox{$N_G^+(|V(G)|-1) = \{|V(G)|-2\}$},~
         \mbox{$|N_G^-(|V(G)|-1)| \geq 2$}, and, 
         for all $v \in V(G) \setminus \{0, |V(G)|-1\}$,
         $N_G^+(v) = \{v-1, w\}$, for some $w > v$.
\end{enumerate}
\end{definition}

From now on, without loss of generality, we shall take $\sigma$ for granted 
and assume the vertex set of any self-labeling
reducible flow graph $G$ \emph{is} the very set $V(G) = \{0, 1, \ldots, |V(G)|-1\}$. 
By doing so, we may simply compare two vertices, e.g. $v > u$ 
(or $v$ greater than $u$, in full writing), 
whereas we would otherwise need to compare their images under $\sigma$, 
e.g. $\sigma(v) > \sigma(u)$. 

\begin{definition}\label{def:representative-tree}
The \emph{representative tree} $T$ of a self-labeling reducible flow graph $G$ 
with Hamiltonian path $H$ has vertex set 
$V(T) = V(G) \setminus \{0\}$ and edge set $E(T) = E(G) \setminus E(H)$, where all edges are deprived of their orientation. 
\end{definition}

\begin{figure}
    \centering
    \includegraphics[width=13cm]{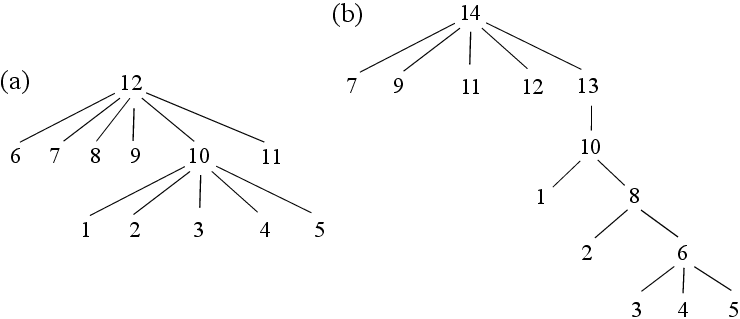}
    \caption{\label{f:trees} Representative trees of the watermark graphs produced by Chroni and Nikolopoulos's encoding algorithm for identifiers (a) $\omega = 31$ and (b) $\omega=43$ (the full watermark for $\omega = 43$ is shown in Figure~\ref{f:graph43}). It is easy to check that such graphs are self-labeling reducible flow graphs.}
\end{figure}

A representative tree $T$ is always regarded as a \emph{rooted} tree whose root is $|V(G)|-1$, 
Moreover, it is regarded as an \emph{ordered} tree, that is, 
for each $v \in V(T)$, 
the children of $v$ are always considered according to 
an ascending order of their labels. 
For $v\in T$, we denote by $N^*_T(v)$ the set of descendants of
$v$ in $T$. 
Figure~\ref{f:trees} depicts two representative trees.

\begin{observation}
The representative tree $T$ 
of a self-labeling reducible flow graph $G$ 
satisfies the max-heap property, that is, 
if vertex $u$ is a child of vertex $v$ in $T$, then $v > u$.
\end{observation}
{\it Proof: }
Direct from the way $T$ is rooted and from property (ii) 
in the definition of self-labeling reducible flow graphs,
whereby the in-neighbors of $w$ in $G \setminus E(H)$
comprise only vertices $v < w$. 
We convey the idea that a representative tree $T$
satisfies the max-heap property by saying that
$T$ is a \emph{descending}, ordered, rooted tree.
$\Box$ \bigskip

\begin{definition}
Let $S = (s_i), i = 1, \ldots, 2n+1$, be a self-inverting permutation.
We say $S$ is \emph{canonical} if:
\begin{enumerate}[label=(\roman*)]
\item there is exactly one $1$-cycle in $S$; 
\item each 2-cycle
$(s_i,s_j)$ of $S$ satisfies $1 \leq i \leq n$, for $s_i > s_j$;
\item  $s_1, \ldots, s_{n+1}$ is a bitonic
subsequence of $S$ starting at $s_1 = n+1$
and ending at $s_{n+1} = 1$. 
\end{enumerate}
\end{definition}

\begin{lemma}\label{lemma:f}
In any canonical self-inverting permutation, 
the fixed element $f$ satisfies $f \in [n+2, 2n+1]$.
\end{lemma}
{\it Proof: }
By property (ii) of canonical self-inverting permutations, 
each $2$-cycle of $S$ must contain at least one element 
whose index $i$ satisfies $1 \leq i \leq n$. 
From property (i), and given the size of $S$, 
it follows that the number of $2$-cycles in $S$ is $n$,
hence, by the pigeonhole principle, 
each and every $2$-cycle in $S$ contains \emph{exactly}
one such element $s_i$ with $1 \leq i \leq n$. 
But this means the other element in each $2$-cycle, 
namely $s_j$, satisfies $s_j \in [n+1, 2n+1]$.
Since there are $n+1$ values in that range and only $n$
such elements $s_j$,  
there must be exactly one element $s_k \in [n+1, 2n+1]$
which is not part of a $2$-cycle, 
and therefore $s_k = f$. 
Now, by property (iii), $n+1 = s_1$,
hence $f \neq n+1$,
and the lemma follows.
$\Box$ \bigskip

Let $T$ be a representative tree.
The {\it preorder
traversal} $P$ of $T$ is a sequence of its vertices that is
recursively defined as follows. If $T$ is empty, $P$ is also empty.
Otherwise, $P$ starts at the root $r$ of $T$, followed by the
preorder traversal of the subtree whose root is the smallest 
child of $r$,
followed by the preorder traversal of the subtree 
whose root is the second smallest child of~$r$, and so on. 
The last (rightmost) element of $P$ is also referred to as
the rightmost element of $T$.

\begin{lemma}\label{l:preorder}
The preorder traversal of a representative tree $T$ is
unique. Conversely, a representative tree $T$ is uniquely
determined by its preorder traversal.
\end{lemma}
{\it Proof: }
We use induction on $|V(T)|$. If $|V(T)| \leq 1$, 
the lemma holds trivially. Let $|V(T)| > 1$, and let $v_k$
be the uniquely defined leaf of $T$ for which 
the path $v_1, \ldots, v_k$ from the root $v_1$ of $T$ to $v_k$
has the property that each $v_i$, for $1 < i \leq k$, 
is the greatest vertex among the children of $v-1$. 
By the induction hypothesis, the preorder traversal $P'$ 
of $T \setminus \{v_k\}$
is unique. 
Because $v_k$ is necessarily the rightmost
vertex of $T$, the preorder traversal $P$ of $T$ 
is uniquely determined as $P'||v_k$.

Conversely, let $P$ be a preorder traversal of some 
representative tree $T$. If $|P| \leq 1$, there is nothing to prove. 
Otherwise, suppose the lemma
holds for preorder traversals of size $\leq k$, and consider $|P|
= k$. Let $v_k$ be the rightmost element of $P$. 
Clearly, $v_k$ must be a leaf of $T$, and also 
the rightmost (i.e.,~greatest) vertex among the children of its parent.
Now define $P' = P - \{v_k\}$. 
By the induction hypothesis, there is a unique tree $T'$ whose preorder
traversal is $P'$. 
Let $v_{k-1}$ be the rightmost element of~$P'$.
We obtain $T$ from $T'$, by making $v_k$ the rightmost child 
of the smallest ancestor $v_j$ of $v_{k-1}$ satisfying $v_j > v_k$,
so $P$ is clearly the preorder traversal of $T$. 
Since no other parent for $v_k$ would be possible
without breaking the ascending order of siblings in a 
representative tree, $T$ is uniquely defined by~$P$.
$\Box$ \bigskip

The first element of the preorder traversal $P$ of a tree $T$ is always its root.
If we remove the first element of $P$, the remaining sequence
is said to be the \emph{root-free} preorder traversal of $T$.

We can now define the class of canonical reducible permutation graphs.

\begin{definition}\label{def:crpg}
A canonical reducible permutation graph $G$
is a self-labeling reducible flow graph on $2n+3$ vertices, for some integer $ \geq 1$,
such that the root-free preorder traversal of the 
representative tree of $G$
is a canonical self-inverting permutation.
\end{definition}

\begin{lemma}\label{l:rpg}
If $G$ is a watermark instance produced by Chroni and Nikolopoulos's 
encoding algorithm~\cite{chroni-2},
then $G$ is a canonical reducible permutation graph.
\end{lemma}
{\it Proof: }
Recall, from Section~\ref{s:nikolopoulos}, that the watermark
graph $G$ associated to identifier $\omega$, 
whose binary representation $B$ has size $n$, is constructed with vertex set 
$V(G) = \{0, \ldots, 2n+2\}$ and an edge set $E(G)$
which can be partitioned into path edges and tree edges
in such a way that all conditions in the definition 
of self-labeling reducible flow graphs are satisfied, 
as can be easily checked. 
Now, by Property~\ref{prop_uk} of Chroni and Nikolopoulos's watermarks (see Section~\ref{s:properties}),
the tree edges of $G$ constitute a representative tree $T$ of $G$
whose root-free preorder traversal is precisely
the self-inverting permutation $P_s$
determined by the encoding algorithm from~\cite{chroni-2}
as a function of $B$.
Consequently,
what is left to prove is that $P_s$ is canonical.
The first condition to $P_s$ being canonical
is asserted by Property~\ref{prop_fixo} in Section~\ref{s:properties} (the fixed element $f$ corresponds to the unique $1$-cycle in $P_s$);
the second condition is given by Property~\ref{prop_n_primeiros_elementos_maiores_que_n};
and, finally,
Properties~\ref{prop_elemento_central_1} and~\ref{prop_inicio_bitonico}
fulfill the third condition, 
therefore $P_s$ is canonical.
$\Box$ \bigskip

\begin{lemma}\label{l:rpg-volta}
If $G$ is a canonical reducible permutation graph, 
then $G$ is the watermark produced
by Chroni and Nikolopoulos's 
encoding algorithm~\cite{chroni-2}
for some integer identifier $\omega$.
\end{lemma}
{\it Proof: }
Let $G$ be a canonical reducible permutation graph,
and $T$ its representative tree.
By Lemma~\ref{l:preorder},
$T$ is uniquely defined by its preorder traversal~$P$.
We show that $P$ corresponds to the 
self-inverting permutation $P_s$ generated by the encoding algorithm of~\cite{chroni-2}
(please refer to Section~\ref{s:nikolopoulos} for details)
when computing the watermark for some integer identifier $\omega$.
By definition,
$P = (s_i), i = 1, \ldots, 2n+1,$ is a canonical self-inverting permutation
presenting a single $1$-cycle $f$
and a number $n$ of $2$-cycles $(p, q)$.
Those $2$-cycles $(p, q)$ define
exactly one bitonic permutation $P_b = (b_j), j = 1, \ldots, 2n+1$
satisfying Property~\ref{prop_end_pb} of Chroni and Nikolopoulos's watermarks with
\begin{enumerate}[label=(\roman*)]
\item $b_{n+1} = f$, and,
\item for all $j \in \{1, \ldots, n\}$, $b_j = p$ if and only if $b_{2n+1-j} = q$.
\end{enumerate}
Such bitonic permutation $P_b$ can be regarded as $Z_0||Z_1^R$
by assigning to $Z_0$ the prefix of $P_b$ comprising its maximal ascending subsequence,
and now the indexes of $0$'s and $1$'s in the extended binary $B^*$
are totally determined. 
We proceed by extracting the binary $B$ that is the one's complement of
the subsequence of $B^*$ with digits 
from the $(n+1)$th to the $(2n)$th position in $B^*$.
Regarding $B$ as the binary representation of a positive integer $\omega$, the
image of such $\omega$ under the encoding function of Chroni and Nikolopoulos 
is isomorphic to $G$.
$\Box$ \bigskip

We proceed to the last definitions before we can give an appropriate, algorithmic-flavored characterization of 
canonical reducible permutation\linebreak  graphs.
%
Let $T$ be the representative tree of some canonical reducible permutation
graph $G$, and $P$ a canonical self-inverting permutation 
corresponding to the root-free preorder traversal of $T$.
We refer to the fixed element $f$ of $P$ also as the fixed element (or vertex) 
of both $G$ and $T$. Similarly, the $2$-cyclic elements of $P$ correspond 
to \emph{cyclic} elements (or vertices) of both $G$ and $T$. 
%
A vertex $v \in V(T) \setminus \{2n+2\}$ is considered \emph{large}
when $n < v \leq 2n+1$; otherwise, $v \leq n$ and $v$ is dubbed as \emph{small}.
Denote by $X,Y$, respectively, the subsets of 
large and small vertices in $T$, 
so $|X| = n+1$ and $|Y| = n$. 
By Lemma~\ref{lemma:f}, $f \in X$. 
We then define $X_c = X \setminus \{f\} = \{x_1, \ldots, x_n\}$
as the set of large cyclic vertices in $T$.  

\begin{figure}
    \centering
    \includegraphics[height=5.4cm]{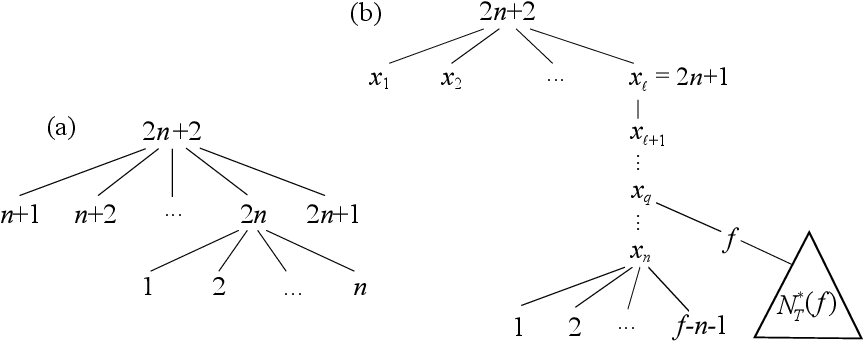}
    \caption{\label{f:types} (a) A Type-1 representative tree. (b) A Type-2 representative tree.}
\end{figure}

\begin{definition}\label{def:T-1}
A representative tree $T$ is a \emph{Type-$1$} tree 
--- see Figure~\ref{f:types}(a) --- 
when
\begin{enumerate}[label=(\roman*)]
\item $n+1, n+2, \ldots, 2n+1$ are children of the root $2n+2$ in $T$; and
\item $1,2, \ldots, n$ are children of $2n$.
\end{enumerate}
\end{definition}

\begin{definition}\label{def:T-2}
A representative tree $T$ is a
\emph{Type-$2$} tree relative to $f$ 
--- see Figure~\ref{f:types}(b) --- 
when
\begin{enumerate}[label=(\roman*)]
\item $n+1 = x_1 < x_2 < \ldots < x_{\ell} = 2n+1$ are the children of $2n+2$, for some $\ell \in [2,n-1]$;
\item $x_i > x_{i+1}$ and $x_i$ is the parent of $x_{i+1}$, for all $i \in [\ell, n-1]$;
\item $1, 2, \ldots, f-n-1$ are children of $x_n$;
\item $x_i = n+i$, for $1 \leq i \leq f-n-1$;
\item $f$ is a child of $x_q$, for some $q \in [\ell, n]$ satisfying $x_{q+1} < f$ whenever $q < n$; and
\item $N_T^*(f) =  \{f-n, f-n+1, \ldots, n\}$ and $y_i \in N_T^*(f)$ has index $x_{y_{i}} - f +1$ in the preorder traversal of $N_T^*[f]$.
\end{enumerate}
\end{definition}

\begin{lemma}\label{lem:rightmost}
If $y_r$ is the rightmost vertex of a
Type-2 representative tree $T$ relative to some $f \neq 2n+1$,
then $y_r$ is equal to the number $\ell$ of children of the root $2n+2$ in $T$.
\end{lemma}

{\it Proof: }
By the definition of a Type-$2$ representative tree,
the only non-leaf child of the root $2n+2$ of $T$
is its rightmost child $x_\ell$, therefore
each child $x_i$ of $2n+2$, for $1 \leq i \leq \ell$,
appears precisely at the $i$th position in 
the root-free preorder traversal $P$ of $T$.
Since, by definition, 
$P$ is self-inverting,
and $y_r$ is the last, $(2n+1)$th
element of $P$, it follows that $y_r$ must be equal to the
index of $2n+1 = x_\ell$ in $P$, that is, $y_r = \ell$.
$\Box$ \bigskip

The following theorem characterizes canonical reducible permutation\linebreak graphs 
in terms of the above defined trees. Such characterization is crucial for the remainder of the paper,
since its straightforward conditions can be checked in linear time
and give rise to the decoding and recovering algorithms that will come next.

\begin{theorem}\label{thm:charac-T}
A digraph $G$ is a 
canonical reducible permutation graph if, and only if,
$G$ is a self-labeling reducible flow graph and
\begin{enumerate}[label=(\roman*)]
\item the fixed element of $G$ is $2n+1$ and $G$ has a Type-$1$
representative tree; or \item the fixed element of $G$ belongs to
$[n+2, 2n]$ and $G$ has a Type-$2$ representative tree.
\end{enumerate}
\end{theorem}
{\it Proof: }
Let $G$ be a canonical reducible permutation graph 
and $T$ its representative tree. 
By definition, $G$ is a self-labeling reducible flow graph.
Let $P = s_1, \ldots, s_{2n+1}$ be the
root-free preorder traversal of $T$, hence
a canonical self-inverting permutation, also by definition. 
This means, among other things, that $P$
has a unique fixed element $f$, and that
$P' = s_1, \ldots, s_{n+1}$ is a bitonic subsequence of $P$. 
Since $T$ is descending, it follows that the prefix
$A$ of $P'$ constituting its maximal ascending subsequence
must comprise solely vertices that are children of the root $2n+2$,
the rightmost of which certainly being $2n+1$.

First, let $f = 2n+1$. Since $f$ constitutes a $1$-cycle of $P$,
$f$ must occupy the rightmost, $(2n+1)$th position in $P$, hence
$f$ is a leaf of $T$. 
Furthermore, by Property~\ref{prop_primeiros_elementos} of 
canonical reducible permutation graphs, 
it follows that $P'$ consists of elements $n+1, n+2, \ldots, 2n, 1$, 
hence $A = n+1, n+2, \ldots, 2n$, and these
vertices are therefore children of $2n+2$ in $T$.
Now, again by Property~\ref{prop_primeiros_elementos},
elements $1, \dots, n$ appear, in this order, to the right of $A$ in $P$.
Considering that $P$ is a preorder traversal and
a representative tree satisfies the max-heap property,
we conclude that vertices $1, \ldots, n$ can only be children of $2n$,
hence $T$ is a Type-$1$ tree, as required.

Next, suppose $f < 2n+1$. By Lemma \ref{lemma:f},
it follows  that $f \in
[n+2, 2n]$. We already know  that the children of $2n+2$ are the
vertices of  $A$. Let $D$ be the subset formed by the remaining
vertices of $P'$. Clearly, the vertices of $D$ must appear in
descending order. Since $T$ satisfies the max-heap property 
and $P$ is a preorder traversal of $T$, 
it follows that the largest vertex of $D$ is a child
of $2n+1$, and subsequently each vertex in $D$ is the parent in
$T$ of the vertex placed to its left along the sequence $D$. 
Again, because $T$ satisfies the max-heap property,
$f \in [2n+2, 2n]$ must be the child of the
smallest vertex $x_q \in D \cup \{2n+1\}$ satisfying $x_q > f$. 
Let us again examine the ascending subsequence $A$. 
We know that the first vertex of $A$ is $n+1$. 
Suppose the leading $k$ vertices of $A$ are $n+1, n+2, \ldots, n_k$, for some $k$. 
Because $P$ is self-inverting, it follows that the vertices $1,2, \ldots, k$ 
must be the children of the last (i.e.,~smallest) vertex of $D$, 
and $k = f-n-1$ by Property~\ref{prop_primeiros_elementos}.
It remains solely to describe how the remaining small
vertices, namely $f-n, f-n+1, \ldots, n$, are placed in $T$. 
Since they appear after $f$ in $P$,
it can only be that this subset comprises exactly the descendants
$N^*_T(f)$ of $f$ in $T$. 
Each of the vertices $y \in N^*_T(f)$
constitute a $2$-cycle with some vertex $x$ belonging to the bitonic
subsequence $P'$, hence the index of $y$ in $P$ is exactly
$x$, and all the conditions for a Type-$2$ tree have thus been verified.

Conversely, let $G$ be a self-labeling reducible flow graph. First,
suppose that (i) applies and let $T$ be the corresponding Type-1
representative tree. Then the root-free preorder traversal $P$ of $T$ is 
\[n+1, n+2, \ldots, 2n, 1, 2, \ldots, n, 2n+1.\] 
Regarding $P$ as a permutation of \{1, \ldots,
2n+1\}, we observe that $2n+1$ is the only fixed vertex on it; for
$1 \leq i \leq n$, each element $n+i$ of $P$ has index $i$, while
$i$ has index $n+i$, and $n+1, n+2, \ldots, 2n, 1$ form a bitonic
subsequence of $P$. Consequently, $P$ is a canonical self-inverting permutation, 
and $G$ is a canonical reducible permutation graph.

Finally, suppose (ii) applies. Let $T$ be the corresponding
Type-$2$ representative tree relative to some $f \in [n+2, 2n]$.
The root-free preorder traversal $P$ of $T$ 
consists of 
\[x_1, \ldots, x_{\ell}, x_{\ell +1}, \ldots, x_q, x_{q+1}, \ldots, x_n, 1, 2,
\ldots, f - n - 1, f, P(N^*_T(f)),\] 
where $x_1 = n+1;  x_{\ell} = 2n+1$;
$x_i = n+i$ for $1 \leq i \leq f-n-1$; 
$x_1, x_2, \ldots, x_n, 1$ is a bitonic subsequence of $P$; 
and $P(N^*_T(f))$ denotes the preorder traversal of the
vertices of $N^*_T(f)$, in which each $y_i \in N^*_T(f)$ has index
$x_{y_{i}} - f +1$. 
Observe that, for $1 \leq i \leq f-n-1$, 
$(n+i,i)$ constitutes a $2$-cycle in $P$. 
Moreover, for $f-n \leq i \leq n$, 
vertex $x_i$ forms a 2-cycle with an
element $y_j \in N^*_T(f)$. 
All conditions have been met, thus $P$ is a canonical self-inverting permutation 
and $G$ is a canonical reducible permutation graph.
$\Box$ \bigskip

\begin{corollary}The recognition of canonical reducible permutation graphs can be achieved in linear time.\end{corollary} \label{cor:recognition}
{\it Proof: }
Direct from Theorem~\ref{thm:charac-T} and from the definitions of self-labeling reducible flow graphs, Type-$1$ and Type-$2$ representative trees, all of whose conditions can be verified in linear time easily.
$\Box$ \bigskip

\section{Linear-time decoding ($k\leq2$ missing edges)}\label{s:newalg}

In this section, we analyze the effects of a distortive attack against a watermark 
(i.e.,~a canonical reducible permutation graph) $G$
from which $k \leq 2$ edges were removed.
Note that the unique Hamiltonian path $H$ of $G$ may have been
destroyed by the attack.
The knowledge of $H$ is crucial for determining the labels of the vertices 
(they range from $2n+2$ to $0$ along $H$). 
Our first task is therefore
to determine whether any path edges are missing from $G$, so we can 
restore $H$ and label the vertices accordingly.

\subsection{Reconstructing the Hamiltonian path}

The algorithm given in pseudocode as Algorithm~\ref{alg:hamilton} 
retrieves the unique Hamiltonian path $H$ of a (possibly damaged) watermark $G'$, that is, a graph isomorphic to a canonical reducible permutation graph $G$ minus $k \leq 2$ edges. It employs two subroutines presented separately: \emph{plug\_next\_subpath} and \emph{validate\_labels}. The algorithm itself is straightforward. It basically builds Hamiltonian path candidates for $G$ (possibly by reinserting some edges) and tests whether the vertex labeling implied by each such candidate satisfies some conditions.
It returns the one and only candidate which passes the test.


\begin{algorithm}[t]

\smallskip
\smallskip

input: a damaged watermark $G'$ with $2n+3$ vertices and \\
\hspace*{36pt}two missing edges\\
output: the unique Hamiltonian path $H$ in the original watermark $G$\\

1. Let $V_0$ be the set of all vertices with degree zero in $G'$. 

\smallskip
\smallskip
\smallskip

2. Let ${\mathcal H}$ be the set of all Hamiltonian path candidates obtained by \\
\hspace*{14pt}a call to \emph{plug\_next\_subpath}$(G',V_0,\emptyset)$

\smallskip
\smallskip
\smallskip

3. \textbf{for each} Hamiltonian path candidate $H \in {\mathcal H}$ \textbf{do} \\
\hspace*{1.2cm}\textbf{if} \emph{validate\_labels}$(G', H)$ \textbf{then}\\
\hspace*{2.0cm}\textbf{return} $H$\\

\caption{\label{alg:hamilton}~~\emph{reconstruct\_Hamiltonian\_path}$(G')$}

\end{algorithm}


The procedure \emph{plug\_next\_subpath}, given as Algorithm~\ref{alg:plugnextsubpath}, takes as input a graph $G'$, a path $Q$ with $V(Q) \cap V(G) = \emptyset$, and an output list ${\mathcal H}$, where (restored) Hamiltonian path candidates of $G'$ will be placed after being concatenated to (the left of) a copy of $Q$. It starts by determining a set $S$ of \emph{subpath heads}. This set comprises every vertex $s \in V(G')$ satisfying $d^+_{G'}(s) = 0$, in case $Q = \emptyset$, or $d^+_{G'}(s) \leq 1$, otherwise. Then it computes the collection of all \emph{maximal backward-unbifurcated paths} of $G'$ reaching $S$ (or $S$-bups). An $S$-\emph{bup} of $G'$ is a path $v_j, v_{j-1}, \ldots, v_1$ such that
\begin{itemize}
\item $v_1 = s$, for some $s \in S$; 
\item $(v_k, v_{k-1}) \in E(G')$, for $2 \leq k \leq j$; and
\item the in-degree of $v_k$ in $G' - \{v_1, \ldots, v_{k-1}\}$ satisfies 
\begin{equation*}
  d^-_{G' - \{v_1, \ldots, v_{k-1}\}}(v_k)=\begin{cases}
    1, & \text{if $1 \leq k \leq j-1$};\\
    0, & \text{if $k = j$}.
  \end{cases}
\end{equation*}
\end{itemize}
In other words, starting from some subpath head $s = v_1$, the procedure builds a directed path $Q$ in backwards fashion by concatenating an in-neighbor of $v_k$ to the left of $v_k$, for $k \geq 1$, whenever the indegree of $v_k$ is $1$ in the graph induced by all vertices which have not yet been incorporated to the path. It carries on iteratively this way until, for some $j$, the indegree of $v_j$ in the aforementioned graph is either zero, whereupon it adds the path so obtained to a list of $S$-bups, or greater than one, whereupon it discards the current path. The rationale behind it is that a backward bifurcation on $v_j$ means there are two vertices, say $u$ and $w$, which have not yet been added to the path, both of which are in-neighbors of $v_j$. Since at most one of them, say $u$, may be the tail of a path edge pointing to $v_j$ in $Q$, the other one, $w$, will be the tail of a tree edge pointing to $v_j$, which is not acceptable since $w$ will be to the left of $v_j$ in the path. Whichever the case, the algorithm starts anew with another subpath head $s \in S$ until all of them have been considered and the list of $S$-bups is fully populated. Finally, it appends each $S$-bup $Q'$, one at each time, to the left of $Q$ (by adding a \emph{plausible path edge} $e \notin E(G')$ from the rightmost vertex in $Q'$ to the leftmost vertex in $Q$) and performs one of two possible actions: 
\begin{itemize}
\item if $V(Q') = V(G')$ (i.e., if $Q'$ is a Hamiltonian path of $G'$), than it adds the new path $Q'||Q$ to the output list ${\mathcal H}$; 
\item otherwise, it makes a recursive call to \emph{plug\_next\_subpath} with parameters $G' - V(Q'), Q'||Q$, and the output list ${\mathcal H}$. 
\end{itemize}
When all $S$-bups have been considered, it returns ${\mathcal H}$.

\begin{algorithm}[t]
\smallskip
\smallskip
input: a graph $G'$, a path $Q$ with $V(Q) \cap V(G') = \emptyset$,\\
\hspace*{36pt}and an output list ${\mathcal H}$\\
output: an updated ${\mathcal H}$ containing all $Q'||Q$\\ 
\hspace*{43pt}where $Q$ is a Hamiltonian path of $G'$ (plus $k \geq 0$ extra edges)\\
\hspace*{43pt}ending at a vertex of degree $d \leq 1$
(or $d = 0$, if $Q$ is empty)\\

1. Let $S \gets \{s \in V(G') \colon d^+_{G'}(s) = 0\}$.\\
\hspace*{14pt}\textbf{if} $Q \neq \emptyset$ \textbf{then} $S \gets S \cup \{s \in V(G) \colon d^+_{G'}(s) = 1\}$

\smallskip
\smallskip
\smallskip

2. \textbf{for each} $s \in S$ \textbf{do}\\
\hspace*{1.2cm}$v \gets s$\\
\hspace*{1.2cm}$Q' \gets s$\\
\hspace*{1.2cm}\textbf{while} $|N^-_{G' - (V(Q') - \{v\})}(v)| = 1$ \textbf{do}\\
\hspace*{2.0cm}$v \gets$ the unique element in $N^-_{G' - (V(Q') - \{v\})}(v)$\\
\hspace*{2.0cm}$Q' \gets v||Q'$\\
\hspace*{1.2cm}\textbf{if} $|Q'| = |V(G')|$ \textbf{then}\\
\hspace*{2.0cm}${\mathcal H} \gets {\mathcal H} \cup \{Q'||Q\}$\\
\hspace*{1.2cm}\textbf{else if} $|N^-_{G' - (V(Q') - \{v\})}(v)| = 0$ \textbf{then}\\
\hspace*{2.0cm}\emph{plug\_next\_subpath}$\left(G' - V(Q'), Q'||Q, {\mathcal H}\right)$\\
\hspace*{1.2cm}\textbf{else} discard $Q'$~~//~~a backward bifurcation was found

\smallskip
\smallskip
\smallskip

\caption{\label{alg:plugnextsubpath}~~\emph{plug\_next\_subpath}$\left(G', 
Q, \mathcal{H} \right)$}

\end{algorithm}

If $H$ is a path, then we indicate the $j$th element of $H$ (from right to left, starting at $j=0$) by $H[j]$.

The second subroutine invoked by Algorithm~\ref{alg:hamilton} is called \emph{validate\_labels}, shown in pseudocode as Algorithm~\ref{alg:validate}. It takes as parameters a watermark $G'$ (with two missing edges) 
and a candidate Hamiltonian path $H$. 
First, it determines the set $H^* = E(H) \setminus E(G')$ of the $k \leq 2$ plausible path edges that were required by $H$. It then checks whether it is possible to obtain a valid canonical reducible permutation graph $G$ through the insertion of $H^*$ and some set of $2-k$ tree edges into $G'$. It does so by testing the following necessary conditions, where $T$ denotes the representative tree of $G$:

\begin{enumerate}[label=(\arabic*)]
\item vertices $H[2n+1]$ and $H[n+1]$ must be in-neighbors of $H[2n+2]$ in~$G$;
\item the out-degree of vertices $H[2n+1], \ldots, H[1]$ must be $2$ in $G$, and the out-degree of $H[2n+2]$ must be $1$;
\item the number of tree edges that would have to be inserted into $G$ so that the two previous conditions are met must not exceed $2-|H^*|$; and, finally,
\item if vertices $H[1]$ and $H[n]$ are not siblings in $T$, then vertex $H[1]$ must be a child of the $n$th descendant of the root $H[2n+2]$ which is a large vertex (i.e., the $n$th descendant of the root, counted right to left, among those whose indexes in $H$ are greater than or equal to $n+1$); moreover, the index in $H$ of the rightmost vertex in the preorder traversal of $T$ must correspond to the number of children of $2n+2$ in $T$.
\end{enumerate}

The first condition above appears in the definition of both Type-$1$ and Type-$2$ representative trees (Definitions~\ref{def:T-1}~and~\ref{def:T-2}), hence its necessity comes directly from Theorem~\ref{thm:charac-T}. The second condition is due to Definition~\ref{def:crpg} and from the second property in the definition of self-labeling reducible flow graphs (Definition~\ref{def:slrfg}). The third condition obviously comes from the fact that $2$ edges were removed from $G$, and $|H^*|$ edges have already been (re-)inserted at this point. Finally, the last condition is due to property (iii) in the definition of Type-$2$ representative trees and to Lemma~\ref{lem:rightmost}. It certainly applies to Type-$2$ representative trees only, which is precisely the case where vertices $H[1]$ and $H[n]$ are not siblings in the representative tree, by definition. We remark that, if $H$ is indeed the unique Hamiltonian path of a canonical reducible permutation graph $G$, then, for all $v \in V(G)$, the canonical label $v$ satisfies $v = H[v]$.

\begin{algorithm}[t]

\smallskip
\smallskip

input: a graph $G'$, with $|V(G')| = 2n+3$,\\
\hspace*{36pt}and a Hamiltonian path candidate $H$, with $|E(H) \setminus E(G')| \leq 2$\\
output: \emph{True}, if the labeling of $V(G')$ implied by $H$ is valid; \emph{False}, otherwise\\

1. Label the vertices of $G'$ in such a way that $H = 2n+2, 2n+1, \ldots, 0$. \\
\hspace*{14pt}Let $H^* \gets E(H) \setminus E(G')$, and insert $H^*$ into $G'$ obtaining $G''$.\\
\hspace*{14pt}Let also $F$ be the forest obtained from $G''$ by the removal of all (path)\\ \hspace*{14pt}edges in $H$, as well as the isolated vertex $0$, and let \emph{missing\_edges} $\gets 0$.

\smallskip
\smallskip
\smallskip

2. \textbf{for each} $v \in \{n+1, 2n+1\}$ \textbf{do}\\
\hspace*{1.2cm}\textbf{if} $(v, 2n+2) \notin E(G'')$ \textbf{then}\\
\hspace*{2.0cm}\textbf{if} $d^+_{G''}(v) = 2$ \textbf{then} \textbf{return} \emph{False}\\
\hspace*{2.0cm}$E(G'') \gets E(G'') \cup \{(v, 2n+2)\}$; \emph{missing\_edges} $\text{+=}~1$\\
\hspace*{14pt}\textbf{for each} $v \in \{1, \ldots, 2n+1\}$ \textbf{do}\\
\hspace*{1.2cm}\textbf{if} $d^+_{G''}(v) < 2$ \textbf{then} \emph{missing\_edges} $\text{+=}~1$\\
\hspace*{14pt}\textbf{if} \emph{missing\_edges} $> 2 - |H^*|$ \textbf{then} \textbf{return} \emph{False}

\smallskip
\smallskip
\smallskip

3. \textbf{if} vertices $1$ and $n$ are siblings in $F$ \textbf{then}\\
\hspace*{1.2cm}Let $r$ be the rightmost vertex in the preorder traversal of $F$.\\ 
\hspace*{1.2cm}\textbf{if} $r > d^-_{G''}(2n+2)~+ $ \emph{missing\_edges} \textbf{then} \textbf{return} \emph{False}\\
\hspace*{1.2cm}Let $x$ be the length of the unique path from $1$ to $2n+2$ in $F$.\\
\hspace*{1.2cm}\textbf{if} $r + x-2 \neq n$ \textbf{then} \textbf{return} \emph{False}

\smallskip
\smallskip
\smallskip

4. \textbf{return} \emph{True}\\

\caption{\label{alg:validate}~~\emph{validate\_labels}$(G', H)$}

\end{algorithm}

\begin{theorem}\label{thm:alg1}Algorithm~\ref{alg:hamilton} correctly retrieves the original, unique Hamiltonian path from a canonical reducible permutation graph on $2n+3$ vertices from which $k \leq 2$ edges were removed. It runs in $O(n)$ time.
\end{theorem}

The proof of Theorem~\ref{thm:alg1} is somewhat involved and unfortunately demands some case analysis. We therefore postpone it until Section~\ref{s:proofthm}.

\subsection{Determining the fixed vertex}

Suppose the watermark $G$ has been attacked, which resulted in a damaged watermark $G'$, where two unknown edges are missing.
Now we shall recognize the fixed vertex of the original watermark, given the damaged one. Getting to know the fixed vertex of $G$ will play a crucial role
in retrieving the missing tree edges and consequently restoring the original identifier $w$ encoded by $G$.

We describe some characterizations that lead to  an efficient computation of the fixed vertex $f$ of $G$. 
Let $T$ be the representative tree of the original watermark $G$. We consider the case where the two edges
that have been removed belong to $T$. Denote by $F$ the forest obtained from $T$
by the removal of two edges. First, we consider the case $f = 2n+1$.

\begin{theorem}\label{thm:charac-1-F-2}
Let $F$ be a forest 
obtained from the representative tree $T$ by removing two edges, 
where $n > 2$. Then $f = 2n+1$
if, and only if,
\begin{enumerate}
\item vertex $2n+1$ is a leaf of $F$; and
\item the $n$ small vertices of $G'$ are children of $2n$ in $F$, 
with the possible exception of at most two of them, in which case they must be isolated vertices.
\end{enumerate}
\end{theorem}
{\it Proof: }
From Theorem \ref{thm:charac-T}, we know that, when $f = 2n+1$, 
$f$ is the rightmost vertex of $T$, hence a leaf of $F$, implying the necessity of condition~(i). 
Again by Theorem \ref{thm:charac-T}, 
the small vertices of $T$ must immediately follow the rightmost cyclic vertex of $T$, 
namely $2n$. 
Since two edges have been deleted from $T$, 
it follows that all small vertices are children of $2n$ in $F$, 
with the possible exception of at most two of them, 
which then became isolated vertices, 
so condition~(ii) is also necessary. 

Conversely, suppose conditions (i) and (ii) hold, and assume $f \neq 2n+1$. 
Then the second case covered by
Theorem~\ref{thm:charac-T} applies for $T$. 
If $f = 2n$, we know from $n > 2$ that vertex $2n+1$ has at least $3$ children in $T$, 
making it impossible for $2n+1$ to become a leaf of $F$ by the removal of only two edges, 
therefore contradicting condition (i).
If, on the other hand, $f < 2n$, then, again by Theorem \ref{thm:charac-T}, 
vertex $2n+1$ cannot have any small children, contradicting condition~(ii). 
Therefore $f > 2n$, implying $f = 2n+1$.
$\Box$ \bigskip

\begin{figure}[t]
\centering
\includegraphics[width=\textwidth]{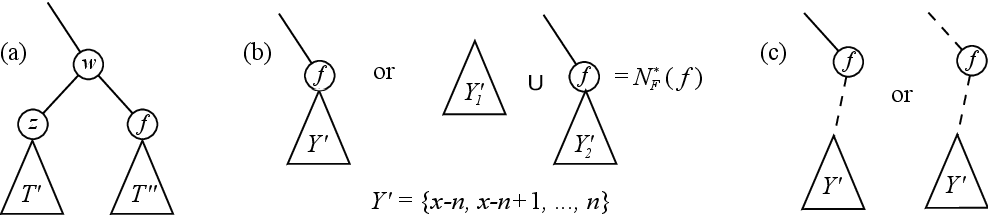}
\caption{\label{f:B-conditions} (a--c) Conditions (i), (ii) and (iii) of Theorem~\ref{thm:charac-2-F-2}, respectively.}
\end{figure}

Next, we characterize the case $f < 2n+1$. Figure~\ref{f:B-conditions} 
helps to visualize the three conditions of the theorem.

\begin{theorem}\label{thm:charac-2-F-2} 
Let $F$ be a forest obtained
from the representative tree $T$ of watermark $G$ by removing two of its edges, 
and let $x \leq 2n$ be a large
vertex of $T$ which is not a child of $2n+2$. Then $x$ is the fixed
vertex $f$ of $G$ if, and only if,
\begin{enumerate}[label=(\roman*)]
\item the large vertex $x$ has a sibling $z$ in $F$,
and $x > z$; or
\item the subset of small vertices $Y' \subset Y$,
$Y' = \{x-n, x-n+1, \ldots, n\}$ can be partitioned into at most
two subsets $Y_1', Y_2'$, such that $\emptyset \neq Y_1' =
N_F^+(x)$ and $Y_2'$ is the vertex set of one of the trees which
form $F$; or, whenever the previous conditions do not hold, 
\item the large vertex $x$ is the rightmost vertex of
one of the trees of $F$, while the rightmost vertices of the
remaining trees are all small vertices.
\end{enumerate}
\end{theorem}

{\it Proof: }
For the sufficiency of condition~(i),
let $x$ be a large vertex of $F$, $z$ a sibling of $x$ in
$F$ and $x_q$ their parent. By Theorem \ref{thm:charac-T}, the
only large vertex of $T$ which is not a child of $2n+2$ and has
some sibling $z$ is precisely the fixed vertex $f$. Clearly, the
removal of edges of $T$ cannot create new vertices having this
property. Furthermore, $x_q \notin \{x_n, 2n+1\}$ implies that $f$ has a
unique sibling $x_{q+1}$, hence $f > x_{q+1}$ according to the
ascending order of siblings in $F$, 
whereas $x_q \in \{x_n, 2n+1\}$ implies every
sibling $y$ of $f$ is a small vertex, hence $f > y$.
Consequently, $x = f$.

Now suppose condition~(ii) holds. First, assume that $Y_2' = \emptyset$.
In this case, $Y' = Y_1' = \{x-n, x-n+1, \ldots, n\} = N_F^*(x)$. Again,
according to Theorem \ref{thm:charac-T}, we can locate a unique
vertex $f$ fulfilling this property, implying $x = f$. In
addition, when $Y_2' \neq \emptyset$, we can again select a unique
vertex $f$, where $N_F^*(f) \cup Y'_2 = Y'$. Thus, $x = f$ indeed.

Finally, assume neither condition~(i) nor condition~(ii) hold. 
Because~(i) is not satisfied,
we have that either $x_q \notin \{x_n, 2n+1\}$, and the
edge from $x_q$ to one of its children has been deleted; or
$x_q \in \{x_n, 2n+1\}$, and the edge $(x_q,f)$ has been deleted.
Additionally, since~(ii) is not satisfied, $f$ must have a unique child $y$, 
and the edge $(f,y)$ has
also been removed. Next, assume that, in such a context, 
condition~(iii) is verified. For the sake of contradiction, suppose the theorem is false, 
so that $x \neq f$. 
Since $x \neq 2n+1$ and $x$ is not a child of $2n+2$, it
follows that it must be a descending vertex, whereupon the fact that $x$ is the
rightmost vertex of the tree of $F$ containing it implies that $x$ is a leaf of $F$. 
Now the latter implies that the
edge $(x_q,x)$ of $T$ has been removed, where $x_q$ is the parent of
$x$ in $T$. Because condition (i) is not satisfied, at
least one edge has been removed from $T$, and because condition~(ii) is not
satisfied, at least one more edge has been deleted from $T$. 
Since no more than two edges overall have been
removed, we conclude that the assumption is false, and therefore,
here again, $x = f$.

Conversely, assume that $x$ is the large vertex of $F$ satisfying
$x = f$. We prove that condition~(i) or condition~(ii) holds, otherwise
condition~(iii) is satisfied.

Let $x_q$ be the parent of $x=f$ in $T$. If $x_q$ has at least two
children in $F$, then $f$ is larger than its
siblings, by Theorem \ref{thm:charac-T}, and condition~(i)
holds. Alternatively, if $f$ is not a leaf of $F$, then the set $Y'
= \{x-n, x-n+1, \ldots, n\}$ either satisfies $Y' = N_F^*(f)$ or it
can be split into two subsets $Y'_1 \cup Y'_2 = Y'$,
where $Y'_1 = N_F^*(f)$ and $Y'_2$ is the vertex set of one of the trees
of $F$. In this situation, condition~(ii) holds. Assume, next, that neither
condition~(i) nor condition~(ii) hold. Then the parent $x_q$ of $f$ in $T$ has at most one
child in $F$, whereas $f$ has no children. The latter implies that $f$ is the
rightmost vertex of the tree of $F$ containing it. Since $x_q \neq f$, we know
that no more than two edges have been deleted from $T$, hence no large
vertex other than $f$ can be a leaf of $F$. Consequently, condition~(iii) holds,
completing the proof.
$\Box$ \bigskip

The above theorems lead to an algorithm that efficiently finds 
the fixed vertex
of watermark $G$ (see Algorithm~\ref{a:finding_f}).
The input is the forest $F$, 
obtained from the representative tree $T$ of $G$ by the removal of two edges. 
First, the algorithm checks whether $f = 2n+1$. 
By Theorem
\ref{thm:charac-1-F-2}, it suffices to verify whether $2n+1$ is a leaf of $F$ and
all small vertices are children of $2n$, except possibly two, which must be isolated vertices.
If this is not the case, then the algorithm proceeds to determining $f$ knowing that $f < 2n+1$. Basically, such task consists in checking conditions~(i), (ii) and (iii) of Theorem~\ref{thm:charac-2-F-2}, which can be done in a straightforward manner.

\begin{algorithm}[t] 

\bigskip

input: a forest $F$ (a representative tree with two missing edges)\\
output: the fixed element $f \leq 2n+1$\\

1. \textbf{if} $F$ contains a large vertex $x$ having a sibling $z$, \textbf{then}\\
\hspace*{1.2cm}\textbf{return} $f := max\{x,z\}$

\smallskip
\smallskip
\smallskip

2. \textbf{for each} large vertex $x$ of $F$ satisfying $N_F(x) \neq \emptyset$ \\
\hspace*{1.2cm}\textbf{for each} small $y \in N_F(x)$\\ 
\hspace*{2.0cm}$Y' \gets \{x-n, x-n+1, \ldots, n\}$\\
\hspace*{2.0cm}\textbf{if} $(N_F^*(x) = Y'$ \textbf{or} $N_F^*(x) \subset Y')$ \textbf{and}\\ 
\hspace*{2.5cm}$(Y' \setminus N_F^+(x)$ is the vertex set of a tree of $F)$ \textbf{then}\\
\hspace*{2.8cm}\textbf{return} $f := x$

\smallskip
\smallskip
\smallskip

3. Find the preorder traversals of the $3$ trees of $F$, and let $f$ be the\\
\hspace*{14pt}unique vertex
that is both large and the 
rightmost element of the\\
\hspace*{14pt}preorder traversal of some tree of $F$.\\
\hspace*{14pt}\textbf{return} $f$\\

\caption{\label{a:finding_f}~~\emph{find\_f}$(F)$}
\end{algorithm}

Steps 1 and 3 can be computed easily in linear time. As for Step 2, observe that
there are at most two large vertices $x$ of $F$ that may satisfy the condition of having only small children.
Consequently, the tests in Step 2 apply to at most two candidates $x$, 
hence the entire algorithm runs in $O(n)$ time.

\subsection{Determining the root's children}

After having identified the fixed vertex of the watermark, 
we are almost in a position to determine the tree edges that have been removed.

Observe that, when $f = 2n+1$, the task is trivial, since, in this case, 
by Theorem~\ref{thm:charac-T}, 
there can be only one canonical reducible permutation graph $G$ relative to $n$.
Such graph is precisely the one with a Type-$1$ representative tree $T$, 
which is unique for each $n > 2$ 
(cf.~Property~\ref{prop_fixo} of canonical reducible permutation graphs, in Section~\ref{s:nikolopoulos}).
By definition, the root-free preorder traversal of a Type-$1$ representative tree, 
when $f = 2n+1$, 
is $n+1, n+2, \ldots, 2n, 1, 2, \ldots, n, 2n+1$.

We therefore want to determine
the children of $2n+2$
restricted to the case where $f < 2n+1$. 
Let $G$ be a watermark, $T$ its representative tree  
and $F$ the forest obtained from $T$ by the removal of two edges.
As usual, $f$ stands for the fixed vertex of $T$, 
$X$ is the set of large vertices other than $2n+2$, 
and $X_c = X \setminus \{f\}$. 
Finally, denote by
$A \subseteq X_c$ the subsets of ascending large cyclic vertices of $T$, 
which we shall refer to simply as the \emph{ascending} vertices,
and denote by $D$ the set $D = X_c \setminus A$ of 
descending large cyclic vertices of $T$, or simply the \emph{descending} vertices. 
Given the forest $F$ and its fixed vertex $f$, 
Algorithm~\ref{alg:ascending} computes the set
$A$, which, as we recall from the proof of Theorem~\ref{thm:charac-T},
corresponds precisely to the children of the root $2n+2$.

\begin{algorithm}[t]
\smallskip
\smallskip
input: a forest $F$ (a representative tree with two missing edges)\\
output: the children $A$ of the root $2n+2$ of the representative tree\\

1. \textbf{if} $F[X_c] \cup {2n+2}$ is connected \textbf{then} \\
\hspace*{1.2cm}\textbf{return} $A := N_F(2n+2)$

\smallskip
\smallskip
\smallskip

2. \textbf{if} $F[X_c] \cup {2n+2}$ contains no isolated vertices \textbf{then} \\
\hspace*{1.2cm}\textbf{return} $A := N_F(2n+2) \cup {2n+1}$

\smallskip
\smallskip
\smallskip

3. \textbf{if} $F[X_c] \cup {2n+2}$ contains two isolated vertices $x,x'$ \textbf{then} \\
\hspace*{1.2cm}\textbf{return} $A := N_F(2n+2) \cup \{x,x'\}$

\smallskip
\smallskip
\smallskip

4. \textbf{if} $F[X_c] \cup {2n+2}$ contains a unique isolated vertex $x$ \textbf{then}\\
\hspace*{1.2cm} \textbf{if} $|N_F^*(f)| = 2n - f + 1 $ \textbf{then} \\
\hspace*{2cm} let $y_r$ be the rightmost vertex of $N_F^*(f)$ \\
\hspace*{2cm} \textbf{if} $|N_F(2n+2)| < y_r$ \textbf{then} \\
\hspace*{2.8cm}\textbf{return} $A := N_F(2n+2) \cup \{x, 2n+1\}$ \\
\hspace*{2cm} \textbf{else} \\
\hspace*{2.8cm}\textbf{return} $A := N_F(2n+2)$ \\
\hspace*{1.2cm} \textbf{else} \\
\hspace*{2cm}\textbf{return} $A := N_F(2n+2) \cup \{x\}$\\


\caption{\label{alg:ascending}~~\emph{find\_ascending\_large\_vertices}$(F)$}

\end{algorithm}

It is easy to conclude that the above algorithm can be implemented in $O(n)$ time. Now we prove its correctness.

\begin{theorem}\label{thm:alg-ascending}
Algorithm~\ref{alg:ascending} correctly computes the set of ascending vertices $A$ of $T$.
\end{theorem}
{\it Proof: }
We follow the different conditions that are checked by the algorithm.
Assume $F[X_c] \cup \{2n+2\}$ is connected. 
Then $N_T(2n+2) = N_F(2n+2$), implying $A = N_F(2n+2)$. 
The algorithm is therefore correct if it terminates at Step $1$.

Assume $F[X_c] \cup \{2n+2\}$ is disconnected, but has no isolated vertices.
Then either $N_F(2n+2) = N_F(2n+2)$ or the edge $(2n+2,2n+1)$ was one of
those that might have been removed from $T$. 
In any of these situations, we can write $A = N_F(2n+2) \cup {2n+1}$, 
implying that the algorithm is also
correct if it terminates at Step $2$.

Assume $F[X_c] \cup \{2n+2\}$ contains two distinct isolated vertices $x,x'$. 
The only possibility is $x,x' \in N_T(2n+2)$. So, the action of
constructing $A$ as the union of $x,x'$ and $N_F(2n+2)$ assures correctness, 
whenever the algorithm terminates at Step $3$.

The last situation is $F[X_c] \cup \{2n+2\}$ containing a unique isolated vertex $x$. 
We consider the following alternatives. If $|N_F^+(f)| = 2n-f+1$,
it implies that $N_T^*(f) = N_F^*(f)$, 
because the set of descendants of $f$ in $T$ comprises exactly 
$y_{f_0}, y_{f_0+1}, \ldots, y_n$, 
The number of such descendants of $f$ is therefore $n-f_0+1$, 
which, by Property~\ref{prop_fixo} of canonical reducible permutation graphs,
is equal to $2n-f+1$.
Now, by Theorem \ref{thm:charac-T}, $|N_T(2n+2)| = y_r$, 
where $y_r$ is the rightmost vertex of $N_F^*(f)$. 
In this situation, $|N_F(2n+2)| < y_r$ implies
that $x$ necessarily belongs to $N_F(2n+2)$. 
In addition, edge $(2n+2,2n+1)$ might also have been deleted from $T$, 
since a single edge deletion suffices to turn $x$ 
into an isolated vertex.
Observe, on the other hand, that isolating a large vertex which is not a child of $2n+2$
requires the removal of at least two edges, provided $n>2$. 
Thus, $A = N_F(2n+2) \cup \{x, 2n+1\}$, and the algorithm is correct. 
In case $|N_F(2n+2)| = y_r$,
we know that $N_T(2n+2) = N_F(2n+2)$, hence $A= N_F(2n+2)$,
ensuring the correctness of the algorithm. 
Finally, when $|N_F^*(f)| \neq 2n-f+ 1$, it means
some edge inside the subtree rooted at $f$ has been deleted from $T$. 
In this case, the isolated vertex $x$ is necessarily a child of $2n+2$ in $T$, implying
$A = N_F(2n+2) \cup \{x\}$, and the algorithm is correct.
$\Box$ \bigskip

\subsection{Retrieving the missing edges}
Once we know the set of ascending vertices, it is simple to restore the entire tree $T$.  
Basically, given sets $A$ and $X_c$, 
we obtain the set of $D$ of descending vertices.
Then, by sorting $A$ and $D$ accordingly, 
we can locate all the large cyclic vertices in $T$, 
using the model given by Theorem~\ref{thm:charac-T}. 
We then place $f$ in $T$, such
that its parent $x_q$ is smallest cyclic vertex that is larger than $f$. 
Finally, we place the small vertices. Vertices $\{1,2, \ldots, f-n-1\}$ are
all children of $x_n$. The remaining small vertices $\{f-n, f-n+1, \ldots, n\}$ are 
descendants of $f$ and their exact position in $T$  can be obtained as follows.
For each $y \in \{f-n, f-n+1, \ldots, n\}$, we find its position in the preorder 
traversal $P$ of $T$ by determining the large vertex $x$ whose
position in the bitonic sequence of the cyclic large vertices is exactly $y$. 
Then $y$ must be the $x$th vertex in the root-free preorder traversal of $T$. 
Finally, the position of $f$ in $P$ is clearly equal to $f$.

The details are given in Algorithm~\ref{alg:preorder}, which computes the preorder traversal $P$ of $T \setminus {2n+2}$.

\begin{algorithm}[t]

\smallskip
\smallskip
input: the fixed vertex $f$, the set $A$ of ascending vertices,\\
\hspace*{36pt}and the set $X_c$ of large cyclic vertices\\
\hspace*{36pt}in a representative tree $T$\\
output: the preorder traversal $P$ of $T$\\

1. Let $D \gets X_c \setminus A.$

\smallskip
\smallskip
\smallskip

2. The initial vertices of $P$ are those of $A$ in ascending order, \\
\hspace*{14pt}followed by those of $D$, in descending order. \\
\hspace*{14pt}Now, subsequently place in $P$ the
small vertices $1, \ldots, f-n-1$, \\
\hspace*{14pt}in this exact order, immediately after the last descending\\
\hspace*{14pt}vertex $x_n \in D$. Then place $f$ as to immediately
follow $f-n-1$.

\smallskip
\smallskip
\smallskip

3. For each small vertex $y \in \{f-n, f-n+1, \ldots, n\}$, \\
\hspace*{14pt}let $P[y]$ be the (large) vertex $x$ whose index in $P$ is $y$, \\
\hspace*{14pt}and place $y$ at position
$x$ in $P$, i.e., satisfying $P[x]=y$.

\smallskip
\smallskip
\smallskip

4. \textbf{return} $P$\\

\caption{\label{alg:preorder}~~\emph{retrieve\_preorder\_traversal}$(T, f, A, X_c)$}
\end{algorithm}

Again, it is straightforward to conclude that Algorithm~\ref{alg:preorder} 
correctly computes the preorder traversal of $T$ in time $O(n)$.
Such procedure assures the complete retrieval of $T$ 
and therefore we are able to restore the watermark $G$ in full.

\subsection{A new decoding algorithm}\label{s:linearalg}
We can now formulate our new decoding algorithm. 
If the input watermark presents $k\leq2$ missing edges,
the algorithm is able to fix it 
prior to running the decoding step.
The decoding step itself is absolutely straightforward, and relies on the following theorem.

\begin{algorithm}[t]
\smallskip
\smallskip
input: a watermark $G$ with $2n+3$ vertices and $0 \leq k \leq 2$ missing edges\\
output: the identifier $\omega$ encoded by $G$\\

1. Let $k \gets |E(G)| - (4n+3)$.

\smallskip
\smallskip
\smallskip

2. If $k > 2$, report the occurrence of $k$ edge removals and halt.

\smallskip
\smallskip
\smallskip

3. If $0 < k \leq 2$, proceed to the reconstitution of the watermark\\
\hspace*{14pt}(see Section~\ref{s:newalg}, Algorithms \ref{alg:hamilton}--\ref{alg:preorder}).

\smallskip
\smallskip
\smallskip

4. Calculate and return the identifier $\omega$ as indicated by Theorem~\ref{thm:decoding}.\\

\caption{\label{alg:decoding}~~\emph{decode}$(G)$}
\end{algorithm}

\begin{theorem}\label{thm:decoding}
Let $\omega$ be a given identifier and $G$ the watermark corresponding to
$\omega$. Let $A = x_1, \ldots, x_{\ell-1}$ be the ascending sequence of
children of $2n+2$, in the representative tree $T$ of $G$, that are different from $2n+1$. 
Then
\[\omega = \sum_{i = 1}^{\ell-1} 2^{2n-x_i}.\]
\end{theorem}
{\it Proof: }
The children of $2n+2$ in $T$ are the vertices $x_i$ of $G$ which
are the tail of some tree edge of $G$ pointing to $2n+2$.
From Property~\ref{prop_uk2} of canonical reducible permutation graphs, 
such vertices $x_i \neq 2n+1$ 
are precisely those satisfying $x_i = n+z_i$, where $z_i$
is the index of a digit $1$ in the binary representation $B$ of $\omega$.
The summation yielding $\omega$ can now be easily checked, 
since the relative value of a digit $1$ placed at position $z_i$
is $2^{n-z_i} = 2^{n-(x_i-n)} = 2^{2n-x_i}$.
$\Box$ \bigskip

As a consequence of the above theorem, 
whenever the input watermark has \emph{not} been tampered with,
the proposed Algorithm~\ref{alg:decoding} is able to retrieve the encoded identifier
in a very simple way. Note that, in this case, it is not even necessary to obtain the representative tree of
the watermark, since the set $A$ can be determined as $A = N^-_G(2n+2)$.

\begin{theorem}\label{thm:alg-decoding}
Algorithm~\ref{alg:decoding} retrieves the correct identifier, encoded 
in a watermark with up to two missing edges, in linear time.
\end{theorem}
{\it Proof: }
Since the final step of the algorithm clearly runs in linear time,
its overall time complexity relies
on the fact that Algorithms \ref{alg:hamilton}--\ref{alg:preorder} run in linear time themselves,
as proved earlier in the text. The correctness of the algorithm follows 
from the fact that those procedures guarantee the reconstruction
of the original watermark when $k\leq2$ edges have been removed,
and from the correctness of Theorem~\ref{thm:decoding}.
$\Box$ \bigskip

\begin{corollary}\label{cor:resilience}
Distortive attacks in the form of $k$ edge modifications (insertions/deletions)
against canonical reducible permutation graphs $G$,
with $|V(G)| = 2n+3$, \mbox{$n>2$}, can be detected in polynomial time, if $k\leq 5$, 
and also recovered from,
if $k\leq 2$. Such bounds are tight.
\end{corollary}
{\it Proof: }
From Theorem~\ref{thm:alg-decoding}, we know that, for $n>2$,
there are no two watermarks $G_1,G_2$, with $|V(G_1)|=|V(G_2)|=2n+3$,
such that $|E(G_1) \setminus E(G_2)| \leq 2$, otherwise it would not always
be possible to recover from the removal of up to two edges.
Thus, for $n>2$, any two canonical permutation graphs $G_1,G_2$
satisfy 
\begin{equation}\label{eq:bound}
|E(G_1) \setminus E(G_2)| = |E(G_2) \setminus E(G_1)| \geq 3,
\end{equation} 
hence
$G_1$ cannot be transformed into $G_2$ by less than $6$ edge modifications. 
Since the class of canonical permutation graphs can be recognized 
in polynomial-time in light of the characterization given in Theorem~\ref{thm:charac-T},
and since any number $k \leq 5$ of edge modifications made to a graph $G$ of the class
produces a graph $G'$ that does not belong to the class, all distortive attacks of such
magnitude ($k \leq 5$) can be detected.
Now, for $k = 2$, we have three possibilities:

\vspace{-0.2cm}
\begin{enumerate}[label=(\roman*)]
\itemsep -0.1 cm
\item 
two edges were removed;
\item 
two edges were inserted; 
\item 
one edge was removed and one edge was inserted.
\end{enumerate}
If case (i) applies, Theorem~\ref{thm:alg-decoding} guarantees that 
the original graph can be successfully restored. 
If case (ii) or case (iii) apply, then a simple algorithm
in which all possible sets of two edge modifications
are attempted against the damaged graph $G'$
suffices to prove that the original graph $G$
can be restored in polynomial time,
since, as we already know, 
exactly one such set shall turn $G'$
into a canonical reducible permutation graph.
The case $k = 1$ is simpler and can be tackled in analogous manner.

It remains to show that such bounds are tight.  
We present a pair of canonical permutation graphs
$G_1, G_2$, with $|V(G_1)| = |V(G_2)| = 2n+3, n > 2$, such
that inequation (\ref{eq:bound})
holds with equality. We remark that there are many such pairs, 
and the following is but an example.
Let $G_1, G_2$ 
be the watermarks relative to identifiers \mbox{$\omega_1 = 8$}, \mbox{$\omega_2 = 9$}, respectively.
Their edge sets are such that
$E(G_1) \setminus \{(2,3), (7,8), (8,9)\} =
 E(G_2) \setminus \{(2,4), (7,9), (8,10)\},$ completing the proof.
$\Box$ \bigskip

\section{Polynomial-time decoding ($k$ missing edges)}\label{s:poly}

The linear-time recognition of the class of canonical reducible permutation graphs, wrapped up in the form of Corollary~\ref{cor:recognition}, allows the construction of a polynomial-time algorithm to recover watermarks which have been deprived of $k$ edges, for arbitrary values of $k$. 
The proposed algorithm is formally robust~\cite{robust}, since it manages to repair a damaged watermark $G'$ whenever such a thing is possible; otherwise, rather than producing a misled result, it shows that $G'$ does not belong to the family of damaged watermarks that can possibly be recovered. As a certificate for this latter case, it outputs two or more watermarks that may become isomorphic to $G'$ through the removal of exactly $k$ of their edges, thus proving that the intended restore is not at all possible.



Let $G$ be a watermark and $G'$ the graph obtained from $G$ when a certain subset of $k$ edges are removed. The idea is simple. The algorithm attempts the addition to $E(G')$ of each and every $k$-subset of non-edges of $G'$, one subset at a time. After each attempt, it checks whether a valid watermark (i.e., a canonical reducible permutation graph) was produced. If, after trying all subsets, only one graph was recognized as such, then the decoding was successful. Otherwise, 
it displays a set containing all watermark candidates.

Since $|V(G')| = 2n+3$ and $|E(G')| = 4n+3-k$, 
the number of $k$-subsets of non-edges of $G'$ is $${{2n+3 \choose 2} - (4n+3-k) \choose  k}=\mathcal{O}(n^{2k}).$$ Thus, considering the effort of running the recognition algorithm for each one of these watermark candidates, the algorithm runs in overall $\mathcal{O}(n^{2k}) \cdot \mathcal{O}(n)=\mathcal{O}(n^{2k+1})$ time. 

The aforementioned formulation considers that all non-edges of $G'$ could be an edge of the original watermark. However, owing to the particular structure of canonical reducible permutation graphs, relatively few among those non-edges do really stand a chance of belonging to $G$. More precisely, every vertex $v$ of $G$ has out-degree at most $2$, hence $v$ must be the tail endpoint of a most $2 - |N^+_{G'}(v)|$ edges. The multiset $M^*$ of all candidates to being the tail of a missing edge has therefore 
\begin{eqnarray*}
|M*| & = & \sum_{v \in V(G')} \left(2 - |N^+_{G'}(v)|\right)\\
        & = & 2 \cdot |V(G')| - \sum_{v \in V(G')} |N^+_{G'}(v)|\\
        & = & 2 \cdot |V(G')| - |E(G')|\\
        & = & 2 \cdot (2n + 3) - (4n+3-k) \\
        & = & k+3
\end{eqnarray*}
elements (not necessarily distinct), and therefore the $k$ missing edge tails may be chosen in 
${|M^*| \choose k} = \mathcal{O}(k^3)$ different ways. For each $k$-subset 
of $M^*$,
the algorithm must choose the head corresponding to each tail, which can be done in $\mathcal{O}(n)$ ways per edge, for an overall $\mathcal{O}(k^3 n^k)$ number of $k$-subsets of non-edges that shall be tentatively added to $G'$. With the $\mathcal{O}(n)$ running time of the recognition algorithm for each such attempt, we complexity of the whole decoding algorithm is an overall $\mathcal{O}(k^3 n^{k+1})$.

As a matter of fact, it is still possible to cut a whole $\mathcal{O}(k^3)$ factor from that asymptotical complexity, if the labels of the vertices are known. To assume that the labels are known is reasonable in many situations, since each vertex corresponds to a block in the CFG of the software, and, by construction, the watermark graph possesses a Hamiltonian path $2n+2, 2n+1, \ldots, 0$ that corresponds, in the CFG, to a chunk of subsequent blocks. If that is the case, then we know the out-degree, in $G$, of all watermark vertices (the tail and the head of the Hamiltonian path have out-degrees $1$ and $0$, respectively; all other vertices have out-degree $2$) and, consequently, the tails of all missing edges. By running the linear-time recognition algorithm on each possible choice of heads, the decoding algorithm has an overall $\mathcal{O}(n^{k+1})$ time complexity.

\section{Proofs of the properties from Section~\ref{s:nikolopoulos}}\label{s:proofs}

We had postponed the proofs of the properties stated in Section~\ref{s:nikolopoulos} to avoid an overhead of technical pages too early in the paper. We now present the full proofs.

\paragraph{Proof of Property~\ref{prop_end_pb}}

When read from right to left,
the $n$ rightmost elements in $P_b$
correspond to the $n$ first elements
in $Z_1$, i.e~the $n$ first indexes, in $B^*$, where a digit $1$ is located.
Since $B^*$ starts with a sequence
of $n$ contiguous $1$'s, the property ensues.
$\Box$ \bigskip

\paragraph{Proof of Property \ref{prop_n_primeiros_elementos_maiores_que_n}}

In $B^{*}$, digits with indexes $1, 2, \ldots,n$
are all $1$, by construction. Since the $n$ rightmost elements in $P_b$ (i.e.,~elements indexed $n + 2 \leq i \leq n^*$ in $P_b$)
correspond to the first $n$ elements in $Y$, and therefore to the first $n$ indexes of $1$'s in $B^*$, those will always be
precisely the elements of set $S = \{1, 2, \ldots, n\}$. In other words, if $s \in S$, then $s$ will have
index $n^* - s + 1 > n + 1$ in $P_b$.
By the time the elements of $P_b$ are gathered together in pairs with views to defining their placement in $P_s$,
element $s$ will be paired with element $q$ whose index is $n^* - (n^* - s + 1) + 1 = s$. Because $s \leq n$, such $q$ clearly does not belong to $S$, hence $q > n$. Now, because $s$ will be assigned index $q$ in $P_s$, the element with index $s$ in $P_s$ will be its pair $q > n$, concluding the proof.
$\Box$ \bigskip

\paragraph{Proof of Property \ref{prop_fixo}}

The bitonic permutation $P_b$ is assembled in such a way that its $(n+1)$th element $f = b_{n+1}$ is either:
\begin{enumerate}[label=(\roman*)]
\item the $(n+1)$th element of $Z_0$, in case $B^*$ has at least $n+1$ digits $0$; or
\item the $(n+1)$th element of $Z_1$, otherwise.
\end{enumerate}

By construction, the number of $0$'s in $B^*$ is one unit greater than the number of $1$'s in $B$. 

If (i) holds, then
$B$ corresponds to an identifier $w$ that is the predecessor of a power of $2$, 
implying all $n$ digits of $B$ are $1$'s. 
If that is the case, then the desired property follows immediately, 
once the $(n+1)$th element of $Z_0$ will be the index of the $(n+1)$th 
--- i.e.,~the last --- 
digit $0$ in $B^*$. 
Such index is, by construction, $n^*$.

If (ii) holds, then
$f$ is the index of the $(n+1)$th digit $1$ in $B^*$. By construction, the $n$ first digits $1$ in $B^*$
occupy positions with indexes $1, \ldots, n$,
and the $(n+1)$th digit $1$ in $B^*$ corresponds
to the first digit $1$ in the one's complement of $B$.
Since that digit has index $f_0$ in the one's complement of $B$,
and there are in $B^*$ exactly $n$ digits to the left of the one's complement of $B$, the property follows.
$\Box$ \bigskip

\paragraph{Proof of Property \ref{prop_primeiros_elementos}}

From the construction of $P_s$ and
Property~\ref{prop_end_pb},
it follows that the elements that occupy
positions with indexes $1, 2, \ldots, n$ in $P_s$
are the first $n$ elements in $P_b$.
It just occurs that the first $n_1 + 1$ numbers in $P_b$ are
the elements of $Z_0$, i.e.,~the indexes of $0$'s in $B^*$.
Now, the last digit in $B^*$ --- the one indexed $n^*$ --- is always a $0$.
Besides that $0$, the other digits $0$ in $B^*$
have indexes $z = n+d$, where each $d$ is the
index of a digit $1$ in $B$
(the original binary representation of the identifier $\omega$).
While the first digit in $B$ is always $1$,
it is also true that:
\begin{enumerate}[label=(\roman*)]
\item the $f_0 - 1$ first digits in $B$
constitute a seamless sequence of $1$'s, in case there is at least one $0$ in $B$; or
\item \emph{all} $n$ digits of $B$ are $1$'s, 
in which case $\omega$ is the predecessor of a power of $2$.
\end{enumerate}
Whichever the case, Property~\ref{prop_fixo}
allows us to state that there is a sequence of
$f - n - 1$ digits $1$ in $B$
starting at the first digit of $B$.
Such sequence will show up, in $B^*$,
starting at index $n + 1$,
in such a way that the $f - n - 1$ first elements of $Z_0$
will be
$n + 1, n + 2, \ldots, n + (f - n - 1) = f - 1$.
Those elements, as we have seen, will be
precisely the first numbers in $P_b$. Because
there are no more than $n$ such elements,
they will be paired against elements
$1, 2, \ldots, f-n-1 \leq n$
(located from the right end of $P_b$ leftwards)
in order to determine their placement in $P_s$,
and the property follows.
$\Box$ \bigskip


\paragraph{Proof of Property \ref{prop_elemento_central_1}}

If the identifier $\omega$ is not the predecessor of a power of $2$,
then its binary representation $B$, whose first digit is always a $1$,
contains some digit $0$. 
In light of this,
Property~\ref{prop_fixo} implies $f \geq n+2$ for all integers $\omega$,
and the first equality now
follows from Property~\ref{prop_primeiros_elementos}.
The second equality is granted by the self-invertibility of $P_s$, whereby $s_j = u \iff s_u = j$.
$\Box$ \bigskip

\paragraph{Proof of Property \ref{prop_2n_plus_1}}

First, note that $f \neq n^*$ corresponds to the case 
where the identifier $\omega$ is not the predecessor of a power of $2$,
i.e.,~$n_1 < n$.
Because the sequence $Z_0$ has exactly $n_1 + 1$ elements, the last of which being the index $n^*$ of the rightmost digit in $B^*$,
element $n^*$ will always be assigned index $n_1 + 1$ in $P_b$.
As we have seen in the proof of Property~\ref{prop_primeiros_elementos},
for $i \leq n$, the $i$th element in $P_b$ will also be the
$i$th element in $P_s$, for it will be paired against element $i$, indexed $n^* - i + 1$ in $P_b$ (due to the starting sequence of
$n$ digits $1$ in $B^*$). That being said, element $n^*$, indexed $n_1 + 1 \leq n$ in $P_b$,
will have index $n_1 + 1$ in $P_s$ as well.
If $f = n^*$, then the definition of $f$ verifies the property trivially.
$\Box$ \bigskip

\paragraph{Proof of Property \ref{prop_inicio_bitonico}}

We employ again the fact, noted for the first time in the proof of Property~\ref{prop_primeiros_elementos},
that the subsequence consisting of the first $n$ elements in $P_s$
and the subsequence consisting of the first $n$ elements in $P_b$
are one and the same.
Since $P_b$ is bitonic, whatever subsequence of $P_b$ is bitonic too, particularly the one containing its first $n$ elements.
By Property~\ref{prop_elemento_central_1}, the central element
$s_{n+1}$ of $P_s$ is always equal to $1$, therefore the bitonic property of the subsequence consisting of the leftmost elements of $P_s$ will not be broken
after its length has grown from $n$ to $n+1$, that is, after element $s_{n+1} = 1$ has been appended to it.
$\Box$ \bigskip

\paragraph{Proof of Property \ref{prop_uk2}}

The first $n_1 + 1$ elements of the bitonic permutation $P_b$
are the elements of $Z_0$, corresponding to the indexes of $0$'s
in the extended binary $B^*$ 
(which consists, we recall, of $n$ digits $1$, followed by
 the one's complement of the binary representation $B$
of the identifier $\omega$ encoded by $G$, followed by a single digit $0$).
Those elements constitute the ascending prefix $A = n+z_1, n+z_2, \ldots, n+z_{n_1}, 2n+1$, 
where, for $i \in \{1, \ldots, n_1\}$, 
$z_i$ is the index of a digit $1$ in $B$.
From the proof of Property~\ref{prop_primeiros_elementos}, 
we know that, 
for $i \leq n$, 
the $i$th element in $P_b$ will also be the
$i$th element in the self-inverting permutation $P_s$.
Since $n_1 \leq n$, we have that the $n_1$ first elements of $P_s$ 
are precisely the $n_1$ first elements of $A$, hence the tree edge 
tailed at each of those elements must point, by construction, to $2n+2$.
It remains to show that no element $u \notin A \cup \{2n+1\}$ is the tail of a tree edge pointing to $2n+2$.
But this comes easily from the fact that, by Property~\ref{prop_2n_plus_1}, 
the $(n_1+1)$th element in $P_s$ is $2n+1$. Since all vertices $u$ with indexes $i > n_1+1$ in $P_s$
are certainly smaller than $2n+1$, they can only be the tail of tree edges pointing to vertices $q(u) \leq 2n+1$, 
and the proof is complete.
 $\Box$ \bigskip

\paragraph{Proof of Property \ref{prop_uk}}

Both items are trivially verified, since, by construction, every tree edge $(u,k) \in G$ is such that either $k > u$ is the element that is closest to $u$ and to the left of $u$ in $P_s$, or $k = 2n+2$.
$\Box$ \bigskip

\section{Proof of Theorem~\ref{thm:alg1}}\label{s:proofthm}

Let $\mathcal{G}_k$ be the set of all canonical reducible permutation graphs with $k$ edges missing. When an element $G'$ of $\mathcal{G}_k$ is the input of \emph{plug\_next\_subpath}\linebreak$(G', \emptyset, \mathcal{H})$, its output is clearly a Hamiltonian path of some graph $G$ such that $V(G) = V(G')$ and $E(G) \ E(G') \leq k$. Thus, when a canonical reducible permutation $G$ minus two edges is passed to Algorithm~\ref{alg:hamilton}, the path $H$ it returns is the Hamiltonian path of some element of $\mathcal{G}_2$. We claim such graph can be no other but $G$.

Let $\hat{H} = 2n+2, 2n+1, \ldots, 0$ be the unique Hamiltonian path of $G$. 
%
We divide the proof in three cases:
\begin{enumerate}[label=(\roman*)]
\item the removed edges were both tree edges of $G$;
\item the removed edges were both path edges of $G$;
\item the removed edges were one tree edge and one path edge of $G$.
\end{enumerate}

\begin{figure}[t]
\centering
\includegraphics[width=0.98\textwidth]{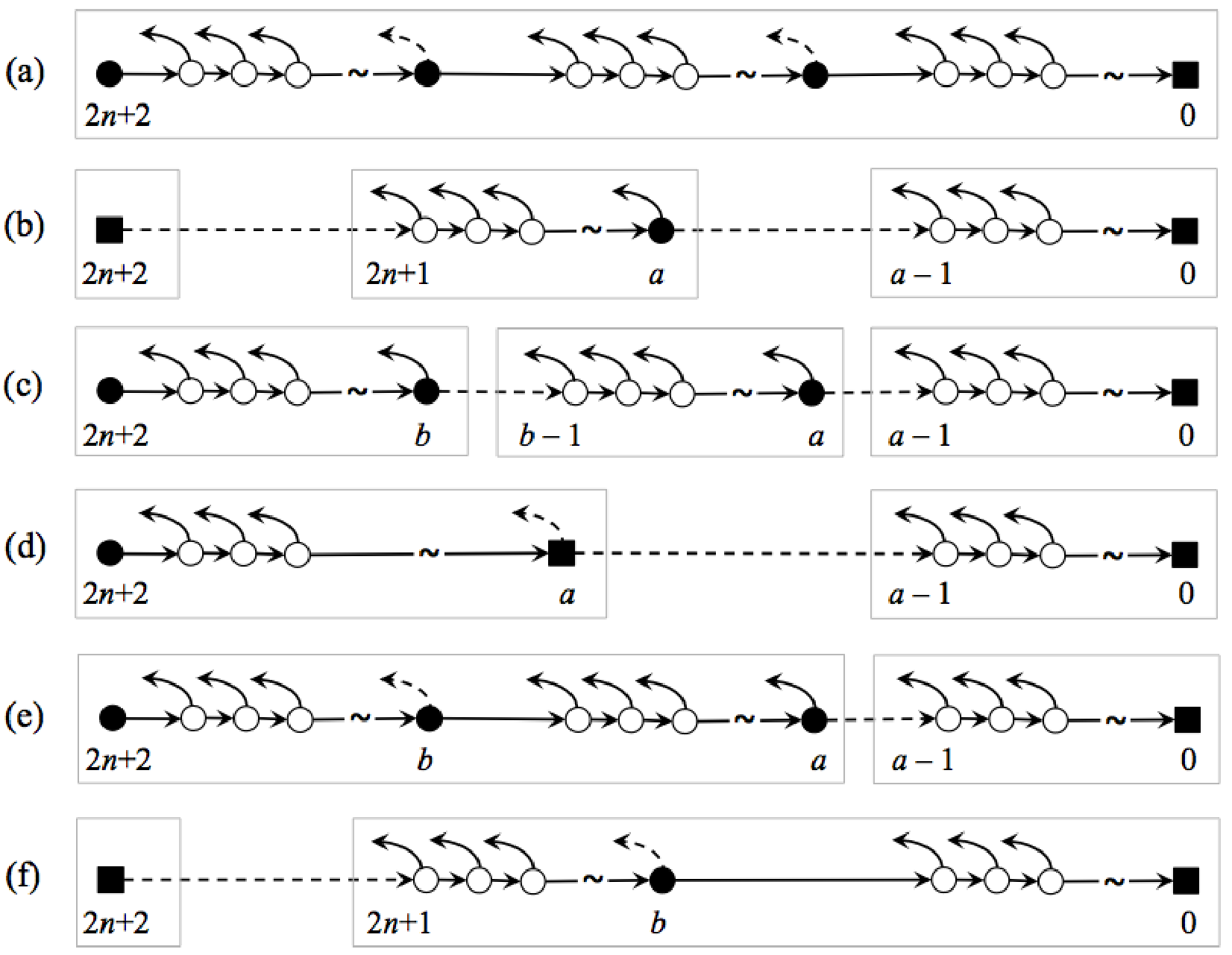}
\caption{\label{fig:cases} Possible scenarios for the Hamiltonian path $H$ of a damaged watermark $G'$. Dashed arrows indicate missing edges. Squares, solid circles and hollow circles represent vertices whose out-degrees in $G'$ are, respectively, $0$, $1$ and $2$. Three hollow circles close together followed by a broken arrow (with a tilde in the middle) indicate subpaths of zero or more edges. Each big rectangle encloses a maximal undamaged subpath of $H$, which corresponds to a maximal backward-unbifurcated path to $s \in V(G)$, i.e., an $\{s\}$-bup.}
\end{figure}

\paragraph{When two tree edges are missing}
The easiest case is (i), as illustrated in Figure~\ref{fig:cases}(a). If only tree edges were removed, then the Hamiltonian path of $G$ is undamaged. Starting from the only vertex with out-degree zero in $G'$, namely vertex $0$, \emph{plug\_next\_subpath}$(G', \emptyset, \mathcal{H})$ outputs $\hat{H}$ at once, never making a single recursive call. Since $\hat{H}$ obviously produces the correct labeling of vertices of $G$, it is validated uneventfully by \emph{validate\_labels}$(G', \hat{H})$ and returned by the algorithm.

\paragraph{When two path edges are missing}
Suppose now that (ii) is the case. Since no tree edges were removed, the only vertex with out-degree zero in $G'$ is vertex $0$, unless the path edge whose tail is $2n+2$ was one of the removed edges (we recall that $2n+2$ has degree $1$ in $G$). We therefore analyze two subcases, according to whether or not $(2n+2,2n+1)$ was removed from $E(G)$.

For the first subcase, suppose the removed path edges were $(2n+2, 2n+1)$ and $(a, a-1)$ for some $a \geq 1$, as in Figure~\ref{fig:cases}(b). Although vertex $2n+2$ has degree zero, its in-degree is greater than $1$ in $G' \setminus \emptyset = G'$, and therefore no $\{2n+2\}$-bup is produced in the main call to \emph{plug\_next\_subpath}. Thus, the only partial path the algorithm produces, starting from vertex $0$ in backwards fashion, is $Q' = a-1, a-2, \ldots, 0$. Because now $a-1$ has no in-neighbors in $G - V(Q')$, it recurses to find possible extensions for $Q'$. The only vertices with degree $0$ or $1$ in $G' \setminus V(Q')$ are now $2n+2$ and $a$. However, any $\{2n+2\}$-bup $Q''$ that may be found will constitute an $H = Q''||Q$ path that will necessarily fail the ensuing validation. This is due to the fact that $H[2n+2]$ will be a vertex other than $\hat{H}[2n+2] = 2n+2$, an out-neighbor of $n+1$ by Theorem~\ref{thm:charac-T}---and the first condition tested by \emph{validate\_labels}$(G', H)$ cannot be met. Because of this, the partial path $Q'$ can only be extended by an $\{a\}$-bup, which can be no other but $2n+1, \ldots, a$, and the remaining vertex $2n+2$ will be concatenated during the next recursive call, completing the Hamiltonian path $2n+2, 2n+1, \ldots, a, a-1, \ldots, 0 = \hat{H}$, as desired.

The second subcase is the one in which the path edge $(2n+2, 2n+1)$ was not removed. Suppose the missing path edges are $(a, a-1)$ and $(b, b-1)$, with $a<b$, as illustrated in Figure~\ref{fig:cases}(b). The first, rightmost subpath located by the algorithm can only be the unique $\{0\}$-bup, namely $Q' = a-1, a-2, \ldots, 0$. Now there are three vertices whose degree are less than or equal to one: $2n+2, b$ and $a$. 

When the algorithm considers $\{2n+2\}$-bups during the recursive call to \emph{plug\_next\_subpath}$(G' - V(Q'), Q'||\emptyset, \mathcal{H})$, whichever ensuing Hamiltonian path candidate $H$ it produces will necessarily be discarded. Indeed, if $n+1 \geq a$, then no bup is even produced because the in-degree of $2n+2$ in $G' - V(Q')$ is at least $2$ by the existence of tree edges $(n+1,2n+2)$ and $(2n+1, 2n+2)$; and, if $n+1 < a$, then $H[n+1]$ is vertex $\hat{H}[n+1] = n+1$ itself and, because its out-degree in $G'$ is already $2$, conditions (1) and (2) checked by \emph{validate\_labels}$(G',H)$ cannot both be met.

When the algorithm considers a $\{b\}$-bup, whichever ensuing Hamiltonian path candidate $H$ it comes up with will also be discarded. Indeed, because the subpath $2n+2, 2n+1, \ldots, b$ of $\hat{H}$ is intact, vertex $2n+2$ will be brought into the $\{b\}$-bup before $v$ does, for all $b-1 \geq v \geq a$, hence $H[2n+2] \neq 2n+2$. Moreover, 
because in particular the tree edge whose tail is $b-1$, say $(b-1, w)$, was not removed, and $w \geq b$, the only possible value for $w$ is $2n+2$, otherwise there would be a vertex $z \in \{2n+1, 2n, \ldots, b\}$ with in-degree greater than $1$ in the subgraph of $G'$ induced by $z$ and by the vertices to the left of $z$ in $H$, which is a contradiction because such path would have been discarded in the last line of Algorithm~\ref{alg:plugnextsubpath}. Thus, vertex $b-1$ is an in-neighbor of $2n+2$ which was not added to the path before $2n+2$ was added. If a backward bifurcation has not arisen, then it is only possible that $b-1$ is precisely the vertex to the left of $2n+2$ in $H$. Repeating the same argument---based on the fact that the tree edge whose tail is $v$ has not been removed---for all $b-2 \geq v \geq a$, we can infer that the only possible Hamiltonian path candidate produced by the concatenation of a $\{b\}$-bup to the left of $Q'$ is $H = a, a+1, \ldots, b-1, 2n+2, 2n+1, \ldots, b, a-1, a-2, \ldots 0$. Now, condition (1) in \emph{validate\_labels} enforces that $H[n+1]$ is the tail of a tree edge pointing to $H[2n+2] = a$. However, because $H[n+1] > a$, such edge cannot be an actual tree edge of the original graph $G$, hence it must be a path edge. Since the only path edge with head $a$ in $G$ is $a+1$, it follows that $H[n+1] = a+1$. And here we shall have a contradiction, since $a+1$ is the second vertex, left to right, in $H$ (i.e., $H[2n+1] = n+1$), unless $a = b-1$. However, if $a = b-1$, then $H[n+1] = b$, and the existence of edge $(b, a) = (H[n+1], H[2n+2])$ is necessary to meet condition (1) in the validation procedure. But $(b, a) = (b, b-1)$ is one of the removed edges, therefore it must be reinserted. Condition (2), on its turn, requires that an outgoing edge is added to $2n+2$ (whose degree is $1$ and whose index $i$ in $H$ satisfies $2n+1 \geq i \geq 1$). Along with the plausible path edge $(b, a-1)$, which was required to concatenate the $\{b\}$-bup to the left of $Q'$, we have a total of $3$ new edges, thus violating condition (3).

Finally, when the algorithm considers $\{a\}$-bups, it necessarily produces the subpath $Q'' = 
b-1, \ldots, a$, which is concatenated to $Q'$, and, because $\{2n+2\}$-bups cannot possibly yield a valid prefix to $Q''||Q'$, the last recursive can only produce the $\{b\}$-bup $2n+2, \ldots, b$, which completes the reconstitution of $\hat{H}$.

\paragraph{When a tree edge and a path edge are missing}

We focus on the the final case (iii), where one path edge and one tree edge were removed. We now consider three subcases separately. In the first one, both the path edge and the tree edge that were removed share the same tail endpoint. In the second one, the tails of the removed edges are distinct. The third case is actually a special case of the second one, when the tail of the removed path edge is vertex $2n+2$.

For the first subcase, illustrated in Figure~\ref{fig:cases}(d), say both removed edges have tail $a \in V(G')$. In this case, vertex $a$ presents degree zero, just like vertex $0$ itself. Any attempts to build a Hamiltonian path $H$ whose suffix is an $\{a\}$-bup, however, shall not succeed. Since vertex $2n+2$ will be brought into $H$ before vertex $0$ does, and because $a$ will be the rightmost vertex in $H$ (i.e., $H[0] = a$), a tree edge leaving $2n+2$ is necessary to satisfy condition~(2) of \emph{validate\_labels}$(G', H)$. But vertex $0$ appears with index $i > 0$ in $H$, and therefore a plausible path edge must be inserted with $0$ as its tail. If the index of $0$ is not $2n+2$, then a tree edge leaving $0$ is also called for. If, on the other hand, the index of $0$ is $2n+2$, then, among the two tree edges reaching $H[2n+2] = 0$ that are required by condition (1) of the validation procedure, at least one of them is still missing. In both cases, condition (3) is violated.

The second case is the one depicted in Figure~\ref{fig:cases}(e), where a path edge $(a, a-1)$, with $1 < a \leq 2n+1$, and a tree edge $(b, v)$, with $v > b$, were removed. Procedure \emph{plug\_next\_subpath} starts by gathering the maximal backward-unbifurcated path $Q'$ whose head is $0$, the only vertex with degree zero in $G'$. The leftmost vertex of such $\{0\}$-bup is vertex $a-1$, the first vertex whose in-degree is zero in the subgraph of $G'$ induced by vertices not in $Q'$, and hence $Q' = a-1, a-2, \ldots, 0$. Now three vertices have out-degree less than or equal to one: $2n+2$, $b$ and $a$.

When the algorithm picks $2n+2$ as a possible continuation of the backward path under construction, the 
index of $2n+2$ in $H$ will be $a$. 
By Theorem~\ref{thm:charac-T}, vertex $2n+1$ is always a child of the root $2n+2$ in the representative tree $T$ of a canonical reducible permutation graph $G$, and, by Property~\ref{prop_uk2}, the number of children $v \leq 2n$ of $2n+2$ in $T$ corresponds to the number $n_1$ of digits $1$ in the binary representation $B$ of the identifier $\omega$ encoded by $G$. As a consequence, the in-degree of $2n+2$ in $G$ is $n_1 + 1$. We now tackle two distinct situations. In the first one, $a \leq n+1$, whereas in the second one $a > n+1$.
If $a \leq n+1$, 
then the in-degree of $2n+2$ in $G'- V(Q')$ is the same as in $G'$ (i.e., $n_1$), since all in-neighbors of $2n+2$ belong to $\{n+1, \ldots, 2n+1\}$ by the same Theorem~\ref{thm:charac-T}. Because, along with the path edge $(a,a-1)$, only one tree edge was removed from $G$ to obtain $G'$, the indegree of $2n+2$ in $G'$ is at least $n_1 + 1 - 1 = n_1$. As a consequence, a backward bifurcation would be noticed on $2n+2$ unless $n_1 = 1$ \emph{and} the tail $b$ of the removed tree edge is one of the in-neighbors of $2n+2$, which in this case are $n+1$ and $2n+1$. 
If $b = n+1$, 
then the tree edge $e = (2n+1, 2n+2)$ is intact, and the only possible placement of vertex $2n+1$ in $H$ is at the position immediately to the left of $2n+2$, so that $e$ functions as a path edge of $H$. Assuming there was no backward bifurcation on $2n+2$ (which would have caused the path $H$ to be discarded), the only possible tree edge leaving $2n$ is $(2n, 2n+1)$, hence $2n$ must be placed to the left of $2n+1$ in $H$. Assuming, similarly, that no backward bifurcation occurred on $2n+1$, the only possible tree edge leaving $2n-1$ is $(2n-1,2n)$, and so on. This reasoning must continue until finally $a$ is concatenated at the very first position of $H$, yielding $H = a, a+1, \ldots, 2n+2, a-1, a-2, \ldots, 0$. Now, condition (1) of the validation procedure requires that $H[n+1]$ and $H[2n+1]$ are in-neighbors of $H[2n+2] = a$. However, this requirement and condition~(2) cannot both be met without violating condition~(3), because, since those two vertices $H[n+1]$ and $H[2n+1]$ are not in $Q'$, they are certainly greater than $a$, but there is only one vertex in $G$ which is greater than $a$ and is an in-neighbor of $a$, namely $a+1$. Therefore an extra tree edge is required, but one extra edge is also required by condition~(2)---a tree edge leaving $b$---and the plausible path edge $(2n+2, a-1)$ had already been inserted, which breaks condition~(3).
We are left with the possibility that the tail of the removed tree edge was 
$b = 2n+1$.
In this case, the tree edge $(n+1, 2n+2)$ is intact, and the vertex immediately to the left of $2n+2$ in $H$ must be $n+1$. Now, since path edge $(n+2, n+1)$ is not the missing one by hypothesis, vertex $n+2$ must be immediately to the left of $n+1$ in $H$, and, since path edge $(n+3, n+2)$ is not the missing one, vertex $n+3$ must appear immediately to the left of $n+2$, and so on, until $b = 2n+1$ is concatenated at the first position of $H$, yielding $H = 2n+1, 2n, \ldots, a, 2n+2, a-1, a-2, \ldots, 0$. To satisfy condition (1) of the validation, vertex $H[n+1]$ must be an in-neighbor of $H[2n+2] = 2n+1$. But, because $n_1 = 1$ ($\omega$ is a power of $2$), the root of its Type-$2$ representative tree has only two children, which allows item (iii) in Definition~\ref{def:T-2} to assure that $2n+1$ has only one child, and this child is not $n+1$, by item (i) of that same definition. Thus, the tree edge $(H[n+1], 2n+1)$ must be added to satisfy condition (1) of \emph{validate\_labels}$(G', H)$, and the only vertices with out-degree $1$ in $G'$ were $b$, which is $2n+1$ itself, $2n+2$, which was already added a plausible path edge connecting it to $a-1$, and $a$. It is therefore only possible that $H[n+1] = a$, that is, the missing path edge is necessarily $(n+1, n)$. And here is where condition (4) of the validation procedure comes into play, enforcing that the root $H[2n+2]$ presents only two children when $\omega$ is a power of $2$. Since that is not the case for the path $H$ so obtained, as can be easily checked, $H$ is discarded. 
The second situation is the one in which 
$a > n+1$. 
This one is easy, since now $H[n+1] = n+1$, which is the tail of a tree edge pointing to $2n+2 \neq H[2n+2]$, and hence conditions (1) and (2) of the validation cannot both be met, unless such tree edge is precisely the one tree edge that was removed. But that would correspond to the subcase shown in Figure~\ref{fig:cases}(d), which we already tackled.

When the algorithm picks $b$ as the head of the first subpath to extend the $\{0\}$-bup $Q'$, all ensuing Hamiltonian path candidates shall be discarded by similar reasons. 

Finally, when it considers the sound continuation $a$, all conditions obviously pass and $\hat{H}$ is delivered.

The third---and last---possible situation is the one depicted in Figure~\ref{fig:cases}(f), where the removed edges were the path edge $(2n+2, 2n+1)$ and a tree edge $(b, v)$, with $v > b$. There are two vertices with degree zero: $0$ and $2n+2$. When the call to \emph{plug\_next\_subpath}$(G', \emptyset, \mathcal{H}\})$ picks $2n+2$ as the rightmost vertex of $Q'$, the leftmost vertex of whatever Hamiltonian path $H$ it produces must be either $0$ or $b$, the only vertices with out-degree less than $2$ in $G'$ (part of the second condition verified by \emph{validate\_labels}). Moreover, the root of the representative tree of $G$ must have only two children (which means $n_1 = 1$, or, equivalently, the identifier $\omega$ encoded by $G$ is a power of $2$), and $b$ must be either $2n+1$ or $n+1$, so that a backward bifurcation does not take place at the very starting vertex $H[0] = 2n+2$. 
If $H[2n+2] = 0$, 
then at least three extra edges are required to put $H$ together and satisfy condition (1) of the validation procedure: a plausible path edge $(H[2n+2], H[2n+1])$, and at least two tree edges, namely $(H[2n+1], H[2n+2])$ and $(H[n+1], H[2n+2])$. But then, of course, condition (3) is violated. 
If $H[2n+2] = b = n+1$, 
then the vertex immediately to the left of $H[0] = 2n+2$ in $H$ must be $H[1] = 2n+1$, and the next vertex right-to-left must be $H[2] = 2n$ and so on, assuming no backward bifurcations took place, until at least vertex $H[n] = n+2$. To put it differently, 
the $\{2n+2\}$-bup $Q'$ considered initially by the algorithm contains (not necessarily properly, depending on whether there was a tree edge pointing to $n+2$ in $G'$) the suffix $Q' = n+2, n+3, \ldots 2n+2$. Now, no matter which vertex $w$ occupies the $(n+1)$th position (right-to-left) in $H$, it was certainly not an in-neighbor of $H[2n+2] = n+1$, because $n+1$ does not have in-neighbors in Type-$2$ trees (and in Type-$1$ trees neither, for that matter). If $w \neq 0$, then $w$ has out-degree $2$, and conditions (1) and (2) of the validation procedure cannot both be met. If, on the other hand, $w = 0$, then $H$ is the concatenation of $Q'$ with the prefix $n+1, n, n-1, \ldots, 0$, an intact subpath of $\hat{H}$. In this case, vertex $H[2n+1]$ is $n$, a vertex with out-degree $2$ in $G'$ which is not an in-neighbor of $H[2n+2] = n+1$, and conditions (1) and (2), again, cannot both be met. 


The verification of the time complexity is straightforward.
$\Box$ \bigskip

\section{Final considerations}\label{s:conclusion}



\begin{figure}
    \centering
    \includegraphics[width=\textwidth]{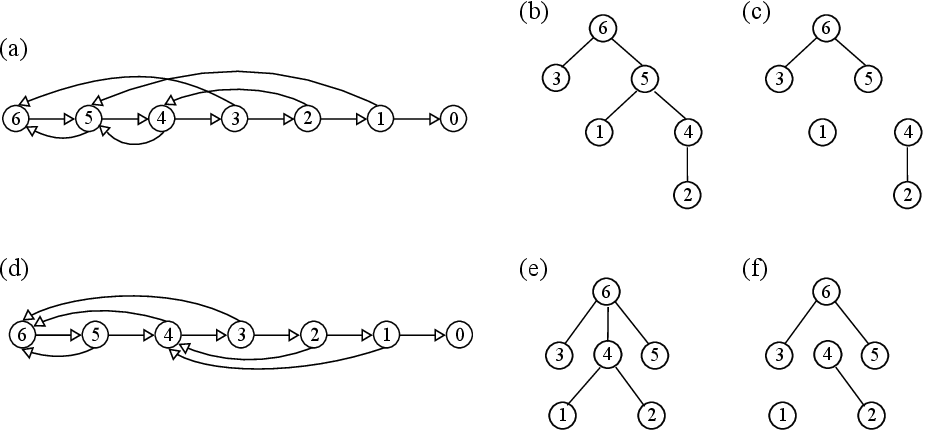}
    \caption{\label{f:n2} (a) The watermark $G_1$ for identifier $\omega=2$; (b) its representative tree $T_1$; (c) the damaged representative tree $T'_1$ obtained from $T_1$ by removing edges $(1,5)$ and $(4,5)$; (d) the watermark $G_2$ for identifier $\omega=3$; (e) its representative tree $T_2$; (f) the damaged representative tree $T'_2$ obtained from $T_2$ by removing edges $(1,4)$ and $(4,6)$. Note that $T'_1$ and $T'_2$ are isomorphic.}
\end{figure}

After characterizing the class of canonical reducible permutation graphs,
we formulated a linear-time algorithm which succeeds in retrieving $n$-bit identifiers  encoded by such graphs (with $n > 2$) 
even if $k \leq 2$ edges are missing. 
Furthermore, we presented 
a polynomial-time algorithm to decode Chroni and Nikolopoulos's watermarks~\cite{chroni-2} with an arbitrary number of missing edges \emph{whenever it is possible to do so deterministically}.

An implication of the first proposed algorithm is that attacks in the form of $k \leq 5$ general
edge modifications (deletions/insertions) can always be \emph{detected} in polynomial time (for $n > 2$), since the replacement of no more than two edges in a canonical reducible permutation graph yields a graph that does not belong to the class. A minimum of six edge modifications (three deletions followed by three insertions) is therefore necessary to change any given watermark into a different, valid watermark.
Indeed, the sole example of two canonical reducible permutation graphs which may become isomorphic to one another when each graph is deprived of only two edges occurs when $n=2$, as illustrated in Figure~\ref{f:n2}. The instances $G_1$ and $G_2$ correspond to 
identifiers $\omega_1=2$ (binary $B=10$) and $\omega_2=3$ (binary $B=11$), respectively. They become isomorphic to one another when edges $(1,5), (4,5)$ are removed from $G_1$ and edges
$(1,4),(4,6)$ are removed from $G_2$. An interesting open
problem is 
to characterize the maximum sets of identifiers $\Omega(k)$ such that, for all $\omega_1, \omega_2 \in \Omega(k)$, the corresponding watermarks cannot become isomorphic to one another when each one is deprived of $k > 2$ edges.




%

Future research focusing on the development
of watermarking schemes resilient to attacks of greater magnitude 
may consider extending the concept of
canonical reducible permutation graphs 
by allowing
permutations with $h$-cycles, with $h > 2$, as well as multiple fixed elements.





%
%
%

\end{document}